\documentclass{article}
\usepackage{epsfig}
\usepackage[hang,nooneline]{subfigure}
\date{\today}

\usepackage{amsmath}
\usepackage{amssymb}

\begin{document}
\title{Glueball condensates as holographic duals of supersymmetric $Q$-balls and boson stars}
\author{{\large Betti Hartmann \footnote{email: b.hartmann@jacobs-university.de} }
and
{\large J\"urgen Riedel \footnote{email: j.riedel@jacobs-university.de} }
\\ \\
{\small School of Engineering and Science, Jacobs University Bremen, 28759 Bremen, Germany}  }

\newcommand{\dd}{\mbox{d}}
\newcommand{\tr}{\mbox{tr}}
\newcommand{\la}{\lambda}
\newcommand{\ka}{\kappa}
\newcommand{\f}{\phi}
\newcommand{\vf}{\varphi}
\newcommand{\F}{\Phi}
\newcommand{\al}{\alpha}
\newcommand{\ga}{\gamma}
\newcommand{\de}{\delta}
\newcommand{\si}{\sigma}
\newcommand{\bomega}{\mbox{\boldmath $\omega$}}
\newcommand{\bsi}{\mbox{\boldmath $\sigma$}}
\newcommand{\bchi}{\mbox{\boldmath $\chi$}}
\newcommand{\bal}{\mbox{\boldmath $\alpha$}}
\newcommand{\bpsi}{\mbox{\boldmath $\psi$}}
\newcommand{\brho}{\mbox{\boldmath $\varrho$}}
\newcommand{\beps}{\mbox{\boldmath $\varepsilon$}}
\newcommand{\bxi}{\mbox{\boldmath $\xi$}}
\newcommand{\bbeta}{\mbox{\boldmath $\beta$}}
\newcommand{\ee}{\end{equation}}
\newcommand{\eea}{\end{eqnarray}}
\newcommand{\be}{\begin{equation}}
\newcommand{\bea}{\begin{eqnarray}}

\newcommand{\ii}{\mbox{i}}
\newcommand{\e}{\mbox{e}}
\newcommand{\pa}{\partial}
\newcommand{\Om}{\Omega}
\newcommand{\vep}{\varepsilon}
\newcommand{\bfph}{{\bf \phi}}
\newcommand{\lm}{\lambda}
\def\theequation{\arabic{equation}}
\renewcommand{\thefootnote}{\fnsymbol{footnote}}
\newcommand{\re}[1]{(\ref{#1})}
\newcommand{\R}{{\rm I \hspace{-0.52ex} R}}
\newcommand{\N}{{\sf N\hspace*{-1.0ex}\rule{0.15ex}%
{1.3ex}\hspace*{1.0ex}}}
\newcommand{\Q}{{\sf Q\hspace*{-1.1ex}\rule{0.15ex}%
{1.5ex}\hspace*{1.1ex}}}
\newcommand{\C}{{\sf C\hspace*{-0.9ex}\rule{0.15ex}%
{1.3ex}\hspace*{0.9ex}}}
\newcommand{\eins}{1\hspace{-0.56ex}{\rm I}}
\renewcommand{\thefootnote}{\arabic{footnote}}

\maketitle

\bigskip

\begin{abstract}
We study non-spinning $Q$-balls and boson stars in 4-dimensional Anti-de Sitter (AdS) space-time.
We use an exponential scalar field potential that appears 
in gauge-mediated supersymmetry (SUSY) breaking in the minimal supersymmetric
extension of the Standard Model (MSSM). We investigate the dependence of the charge and mass of these
non-topological solitons on the negative cosmological constant, the frequency that appears
in the periodic time-dependence as well as on the ratio between the SUSY breaking scale and the Planck mass. 
Next to fundamental solutions without nodes in the scalar
field function we also construct radially
excited solutions. In the second part of the paper we put the emphasis on the
holographic interpretation of these solutions in terms of Bose-Einstein condensates of scalar glueballs 
that are described by a
strongly coupled Quantum Field Theory (QFT) on the boundary of global AdS.
\end{abstract}

\section{Introduction}
Solitons play an important role in many areas of physics. As classical solutions of non-linear field theories, they
are localized structures with finite energy, which are globally regular.
In general, one can distinguish between topological and non-topological solitons.
While topological solitons \cite{ms} possess a conserved quantity, the topological charge, that stems (in most
cases) from the spontaneous symmetry breaking of the theory, non-topological solitons \cite{fls,lp} have a conserved Noether
charge that results from a symmetry of the Lagrangian. The standard example of  non-topological solitons
are $Q$-balls \cite{coleman}, which are solutions of theories with self-interacting complex scalar fields. These objects are stationary with an explicitly
time-dependent phase. The conserved Noether charge 
$Q$
is then related to the global phase invariance of the theory and is directly proportional
to the frequency. $Q$ can e.g. be interpreted as particle number \cite{fls}. 
While in standard scalar
field theories, it was shown
that a non-normalizable $\Phi^6$-potential is necessary \cite{vw}, supersymmetric extensions of the
Standard Model (SM)  also possess $Q$-ball solutions \cite{kusenko}. In the latter case, several scalar fields
interact via complicated potentials. It was shown that cubic interaction terms that result from
Yukawa couplings in the superpotential and supersymmetry (SUSY) breaking terms lead to the existence of $Q$-balls
with non-vanishing baryon or lepton number or electric charge. These supersymmetric
$Q$-balls have been considered as possible candidates for baryonic dark matter 
\cite{dm} and their astrophysical implications have been discussed \cite{implications}.
In \cite{cr}, these objects have been constructed numerically using 
the exact form of a scalar potential that results from gauge-mediated SUSY breaking. However, this
potential is non-differentiable at the SUSY breaking scale.
In \cite{ct} a differentiable approximation of this potential was suggested and the
properties of the corresponding $Q$-balls have been investigated.  
$Q$-ball solutions in $3+1$ dimensions have been studied in detail in 
\cite{vw,kk1,kk2} using a $\Phi^6$-potential.
It was realized
that next to non-spinning $Q$-balls, which are spherically symmetric, spinning solutions
exist. These are axially symmetric with energy density of toroidal shape
and angular momentum $J=mQ$, where $Q$ is the Noether charge of the solution
and $m\in \mathbb{Z}$ corresponds to the winding around the $z$-axis. 
Approximated  solutions of the non-linear partial differential equations
were constructed in \cite{vw} by means of a truncated series in the spherical harmonics to describe
the angular part of the solutions. 
The full  partial differential equation was solved numerically in \cite{kk1,kk2,bh}. 
It was also realized in \cite{vw} that in each $m$-sector, parity-even ($P=+1$)
and parity-odd ($P=-1$) solutions exist. Parity-even and parity-odd 
refers to the fact that
the solution is symmetric and anti-symmetric, respectively with respect
to a reflection through the $x$-$y$-plane, i.e. under $\theta\rightarrow \pi-\theta$.
Complex scalar field models coupled to gravity possess so-called ``boson star'' solutions \cite{kaup,misch,flp,jetzler,new1,new2}.
In \cite{kk1,kk2,bh2} boson stars have
been considered that have flat space-time limits in the form of
$Q$-balls. These boson stars are hence self-gravitating $Q$-balls.

The gravity--gauge theory duality \cite{ggdual} has attracted a lot of attention in the past years. The most famous
example is the AdS/CFT correspondence \cite{adscft} which states that a gravity theory in a $d$-dimensional
Anti-de Sitter (AdS) space--time is equivalent to a Conformal Field Theory (CFT) on the $(d-1)$-dimensional boundary of AdS.
Recently, this theory has been used to describe so-called holographic conductor/superconductor phase transitions
with the help of black holes in higher dimensional space--time \cite{gubser,hhh,reviews}.  
Close to the horizon of the black hole the effective mass of the scalar field can become
negative with masses below the Breitenlohner--Freedman bound \cite{bf} such that the scalar
field becomes unstable and possesses a non--vanishing value on and close to the horizon
of the black hole.
Interestingly, the model used to describe conductor/superconductor phase transitions
by using black holes in higher-dimensional AdS can be modified to describe insulator/superconductor phase
transitions by considering soliton solutions in higher-dimensional AdS. 
This was done in \cite{nrt} and \cite{horowitz}
in (4+1)-dimensional AdS in the probe limit and taking backreaction into account, respectively as well
as in (3+1)-dimensional AdS in \cite{hartmann_brihaye2}.
The AdS soliton is related to the black hole by a double Wick rotation. Moreover, one of the coordinates 
is compactified to a circle.
The AdS soliton has been used before in the context of the description of a confining vacuum 
in the dual gauge theory \cite{witten2,horowitz_myers} since it possesses a mass gap. While for spherically symmetric
black holes in AdS there is the Hawking-Page phase transition from the AdS black hole to global AdS space-time when
lowering the temperature \cite{hawking_page}, this is different for AdS black holes with Ricci-flat horizons 
used in the description of holographic superconductors.
In that case, there is a phase transition between the AdS black hole and the AdS soliton \cite{ssw} which was interpreted as
a confining/deconfining phase transition in the dual gauge theory. Now taking the viewpoint of  condensed
matter this phase transition describes a 1st order insulator/conductor phase transition. 
To complete the picture it was realized in \cite{nrt} that one can add a chemical potential to the AdS soliton.
Increasing the chemical potential $\mu$ the AdS soliton becomes unstable to the formation of scalar hair above some
critical value $\mu_{cr}$. This was interpreted as a 2nd order insulator/superconductor phase transition that 
is possible
even at zero temperature. The study done in the probe limit in \cite{nrt} was extended to take backreaction 
into account \cite{horowitz}. 
Phase diagrams were constructed and it was shown that there is a new type of
phase transition from a superconductor to an insulator for sufficiently strong backreaction. Qualitatively
similar results exist when describing insulator/superfluid phase transitions with non-vanishing
superfluid velocity \cite{hartmann_brihaye2}. Charged solitons in global AdS have also recently been studied in 4 dimensions
\cite{Gentle:2011kv} as well as in 5 dimensions \cite{diasetal,hartmann_brihaye2012}.
  
The limit of vanishing gauge coupling has also been studied in \cite{horowitz}. In that case, the soliton
solutions correspond to planar boson stars in AdS space-time. Since the scalar field
is uncharged the interpretation in terms of insulators/superconductors is difficult in this
case. However, since the AdS/CFT correspondence connects strongly coupled
CFTs to weakly coupled gravity theories the prototype example of a strongly coupled
field theory comes to mind - Quantum Chromodynamics (QCD). As such the planar boson stars in AdS
have been interpreted
as Bose-Einstein condensates of glueballs. Glueballs are color-neutral bound states of gluons predicted by QCD
and the scalar glueball (which is also the lightest possible glueball) is predicted to have a mass of 1-2 GeV (see e.g.
\cite{glueballs_experiment} for an overview on experimental results). Since these glueballs
appear due to non-linear interactions and as such cannot be described by a perturbative approach
it is very difficult to make predictions within the framework of Quantum Field Theory.  
However, holographic methods
have been applied to make predictions about glueball spectra (see e.g. \cite{physics_glueballs} and 
reference therein).

The spherically symmetric
counterparts of boson stars in AdS space-time have been studied before using a massive
scalar field without self-interaction \cite{radu}. In this paper, we will extend the results of \cite{radu} by including a self-interaction potential
of exponential form motivated from supersymmetric extensions of the Standard Model \cite{ct}.
In contrast to the study in \cite{radu} our solutions exist also in the limit of vanishing
gravitational coupling and correspond to $Q$-balls in an AdS background. Moreover, in \cite{radu}
the frequency appearing in the periodic time-dependence of the solutions has been
scaled to unity. In this paper, we will explicitly keep this parameter.

Our paper is organised as follows: in Section 2, we give the model, the Ansatz and the equations
of motion. In Section 3 and 4, respectively, we discuss $Q$-balls in an AdS background and boson stars
in AdS space-time. In Section 5, we give the holographic interpretation of our solutions, while
we conclude in Section 6.

\section{The model}

The existence of $Q$-balls in supersymmetric extensions of the Standard Model has been
suggested some time ago. 
In \cite{cr} these $Q$-balls have been constructed using the following potential
\begin{equation}
\label{real_potential}
 U_{\rm SUSY}(\vert\Phi\vert) = \left\{ 
\begin{array}{l l}
  m^2 \vert\Phi\vert^2  \quad \mbox{if $\vert\Phi\vert \leq \eta_{\rm susy}$ }\\
  m^2 \eta_{\rm susy}^2=const.  \quad \mbox{if $\vert\Phi\vert > \eta_{\rm susy}$}\\ \end{array} \right.
\end{equation}
where $\eta_{\rm susy}$ corresponds to the scale below which supersymmetry is broken and
is roughly $1$ TeV, while $m$ denotes the scalar boson mass.  This potential is not differentiable at $\vert\Phi\vert=\eta_{\rm susy}$. This is
why the following approximation of the above potential has been suggested \cite{ct}
\begin{equation}
\label{potential}
 U(\vert\Phi\vert)=m^2\eta_{\rm susy}^2 \left(1-\exp\left(-\frac{\vert\Phi\vert^2}{\eta_{\rm susy}^2}\right)\right) \ .
\end{equation}
In the following we will study non-spinning $Q$-balls and boson stars in a model containing 
this potential in asymptotically Anti-de Sitter (AdS) space-time.
The action $S$ reads:
\begin{equation}
 S=\int \sqrt{-g} d^4 x \left( \frac{R-2\Lambda}{16\pi G} + {\cal L}_{m}\right)
\end{equation}
where $R$ is the Ricci scalar, $G$ denotes Newton's constant, $\Lambda$ is the negative cosmological constant
with $\ell=\sqrt{-3/\Lambda}$ the Anti-de Sitter radius and ${\cal L}_{m}$ is
the matter Lagrangian given by
\begin{equation}
\label{lag}
 {\cal L}_{m}=-\partial_{\mu} \Phi \partial^{\mu} \Phi^*
 - U(\vert\Phi\vert)
\end{equation}
where $\Phi$ denotes a complex scalar field and we choose as signature of the metric
$(-+++)$. $U(\vert\Phi\vert)$ is the potential given in (\ref{potential}).
The matter Lagrangian ${\cal L}_{m}$ (\ref{lag}) is invariant under the global U(1) transformation
\begin{equation}
 \Phi \rightarrow \Phi e^{i\chi} \ \ \  .
\end{equation}
As such the locally conserved Noether
current $j^{\mu}$, $\mu=0,1,2,3$, associated to this symmetry is given by
\begin{equation}
j^{\mu}
 = -i \left(\Phi^* \partial^{\mu} \Phi - \Phi \partial^{\mu} \Phi^*\right) \  \ {\rm with} \ \ \
j^{\mu}_{; \mu}=0  \ .
\end{equation}
The globally conserved Noether charge $Q$ of the system then reads
\begin{equation}
 Q= -\int \sqrt{-g} j^0 d^3 x  \  .
\end{equation}
Finally, the energy-momentum tensor is given by
\begin{eqnarray}
\label{em}
T_{\mu\nu}&=& g_{\mu\nu} {\cal L} - 2\frac{\partial {\cal L}}{\partial g^{\mu\nu}}\nonumber\\
&=& -g_{\mu\nu} \left[\frac{1}{2} g^{\sigma\rho} 
\left(\partial_{\sigma} \Phi^* \partial_{\rho} \Phi +
\partial_{\rho} \Phi^* \partial_{\sigma} \Phi\right) + U(\Phi)\right] +
\partial_{\mu} \Phi^* \partial_{\nu} \Phi + \partial_{\nu}\Phi^* \partial_{\mu} \Phi\ .
\end{eqnarray}

\subsection{Ansatz and Equations}
For the metric we use the following Ansatz in spherical Schwarzschild-like coordinates
\begin{equation}
 ds^2 = - A^2(r)N(r)dt^2 + \frac{1}{N(r)}dr^2 + r^2 \left(d\theta^2 + \sin^2 \theta d\varphi^2 \right)   \ ,
\end{equation}
where 
\begin{equation}
 N(r)=1-\frac{2n(r)}{r}-\frac{\Lambda}{3} r^2  \ .
\end{equation}
The non-vanishing components of the Einstein tensor in this space-time read
\begin{equation}
 G_{tt}= \frac{A^2 N (1-N - r N')}{r^2}=\frac{A^2 N (2 n' + \Lambda r^2)}{r^2} \ ,
\end{equation}
\begin{equation}
 G_{rr}= \frac{rA N' + 2r A' N + AN - A}{r^2 A N} \ ,
\end{equation}
\begin{equation}
 G_{\theta\theta}= \frac{r}{2A}\left(r A N'' + 3 r A' N' + 2 r A'' N + 2 A N' + 2 A' N\right)
\end{equation}
and $G_{\varphi\varphi}=\sin^2\theta G_{\theta\theta}$. 

For the complex scalar field, we use a stationary Ansatz that contains a periodic dependence of the time-coordinate $t$:
\begin{equation}
\label{ansatz1}
\Phi(t,r)=e^{i\omega t} \phi(r) \ ,
\end{equation}
where $\omega$ is a constant and denotes the frequency.

The coupled system of ordinary differential equations is then given by the Einstein
equations
\begin{equation}
\label{einstein}
 G_{\mu\nu}+\Lambda g_{\mu\nu}=8\pi G T_{\mu\nu}
\end{equation}
with $T_{\mu\nu}$ given by (\ref{em}) and the Klein-Gordon equation
\begin{equation}
\label{KG}
 \left(\square - \frac{\partial U}{\partial \vert\Phi\vert^2} \right)\Phi=0 \ \   \ \ .
\end{equation}
In order to be able to use dimensionless quantities we introduce the following rescalings
\begin{equation}
\label{rescale}
 r\rightarrow \frac{r}{m} \ \ , \ \ \omega \rightarrow m\omega \ \ , \ \  \ell \rightarrow \ell/m \ \ ,
 \  \phi\rightarrow \eta_{\rm susy} \phi \ \ , \ \ n\rightarrow n/m  
\end{equation}
and find that the equations depend only on the dimensionless coupling constant
\begin{equation}
 \kappa=8\pi G\eta_{\rm susy}^2 = 8\pi \frac{\eta_{\rm susy}^2}{M_{\rm pl}^2}  \ ,
\end{equation}
where $M_{\rm pl}$ is the Planck mass. Note 
that $\kappa$ denotes the ratio between
the SUSY breaking scale and the Planck mass which we expect to be small $\kappa\sim 10^{-31}$
if the SUSY breaking scale is on the order of $1$ TeV. In this paper, we will however also study
larger values of $\kappa$ in order to understand the qualitative behaviour of the solutions
in curved space-time. Note that with these rescalings the scalar boson mass $m_{\rm B}\equiv m$ 
becomes equal to unity. 

The coupled ordinary differential equations then read 
\begin{equation}
\label{eq1}
 n' =\frac{\kappa}{2} r^2 \left(N (\phi')^2 + \frac{\omega^2\phi^2}{A^2 N} + 1 - \exp(-\phi^2)\right)  \ ,
\end{equation}
\begin{equation}
\label{eq2}
 A' = \kappa r \left(\frac{\omega^2 \phi^2}{N^2 A} + A \phi'^2\right)  \ ,
\end{equation}
\begin{equation}
\label{eq3}
 \left(r^2 A N \phi'\right)' = -\frac{\omega^2 r^2}{AN} \phi + r^2 A \phi \exp(-\phi^2)  \ ,
\end{equation}
where the prime now and in the following denotes the derivative with respect to $r$.
Note that (\ref{eq1}) corresponds to the $tt$-component of the Einstein equation, while
(\ref{eq2}) is a combination of the $tt$ and $rr$ component:
$g^{tt} G_{tt} - g^{rr} G_{rr} = 8\pi G \left(g^{tt} T_{tt} - g^{rr} T_{rr}\right)$.

These equations have to be solved numerically subject to appropriate boundary conditions.
We want to construct globally regular solutions with finite energy. At the origin we hence require
\begin{equation}
\phi'(0)=0 \ \ , \ \ \ n(0)=0 \ ,
\end{equation}
while we choose
$A(\infty)=1$ (any other choice would just result in a rescaling of the time coordinate). 
Moreover, while the scalar field function falls of exponentially for $\Lambda=0$ with
\begin{equation}
 \phi(r >> 1)=\frac{1}{r} \exp(-\sqrt{1-\omega^2} r) + ...
\end{equation}
it falls of power-law for
$\Lambda < 0$ with
\begin{equation}
 \phi(r>>1)=\phi_{\Delta} r^{\Delta} + ... \ \ , \ \ \Delta=-\frac{3}{2}-\sqrt{\frac{9}{4}+\ell^2}  \ ,
\end{equation}
where $\phi_{\Delta}$ is a constant that has to be determined numerically and
which can be interpreted via AdS/CFT as the value of the condensate of glueballs in the dual theory living
on the boundary of global AdS.

The Noether charge $Q$ of the solutions is given by the following expression
\begin{equation}
 Q= 8\pi \int\limits_0^{\infty} dr \frac{r^2 \omega}{AN} \phi^2   \ ,
\end{equation}
which is well finite also in asymptotically AdS space-time.

For $\kappa\neq 0$, we can read of the mass $M$ from the behaviour of the metric function $n(r)$ at infinity. This reads \cite{radu}
\begin{equation}
 n(r \gg 1)= M + n_1 r^{2\Delta+3} + .... \ , 
\end{equation}
where $n_1$ is a constant that depends on $\ell$. 

For $\kappa=0$ we have $n(r)\equiv 0$, $A(r)\equiv 1$ and 
the mass of the solution is given by the volume integral of the energy-density, which is
given by $T_{00}$:
\begin{equation}
 M=\int d^3 x T_{00}  \ . 
\end{equation}
In our case this gives
\begin{equation}
 M=4\pi \int\limits_0^{\infty} \left[\omega^2 \phi^2 + N^2 (\phi')^2 + N U(\phi)\right] r^2 dr \ ,
\end{equation} 
which is again a finite quantity in both asymptotically flat as well as asymptotically AdS space-time.

\section{$Q$-balls in AdS space-time}
For $\kappa=0$ we find that $A(r)\equiv 1$ and $n(r)=0$, hence $N(r)=1+r^2/\ell^2$. In this case, we are studying
a $Q$-ball in the background of a global AdS space-time. Hence, the space-time
does not backreact. The equation for the scalar field
then reads
\begin{equation}
\label{eqnew}
 \phi''=-\frac{2}{r}\phi' - \frac{N'}{N}\phi' - \frac{\omega^2}{N^2} \phi + 
\frac{\phi \exp(-\phi^2)}{N} \ .
\end{equation}
We can  rewrite the last term in (\ref{eqnew}) as $-\frac{1}{2} dU(\phi)/d\phi$.
and integrate over $\phi$. We then find
\begin{equation}
 \frac{1}{2}\phi'^2 + \frac{1}{2} \frac{\omega^2 \phi^2}{N^2} - \frac{U(\phi)}{2N} = 
{\cal E} - \int\limits_0^r \left(\frac{2}{r}+\frac{N'}{N}\right)(\phi')^2 dr \ .
\end{equation}
The last term is a friction term, ${\cal E}$ is an integration constant. In the classical
mechanics picture adapted e.g. in \cite{vw}, where $\phi$ is considered to describe the
position of the particle and $r$ the ``time-coordinate'', this describes a particle 
with total energy ${\cal E}$ subject to friction moving in a ``time''-dependent effective potential given by
\begin{equation}
 V(\phi)= \frac{1}{2} \frac{\omega^2 \phi^2}{N^2} - \frac{U(\phi)}{2N}  \  \ , \ \ U(\phi)=1-\exp(-\phi^2)   \ .
\end{equation}
Note that this potential becomes ``time''-independent for $G=0$ and $\Lambda=0$, i.e. for
$Q$-balls in a fixed Minkowski space-time since $N(r)\equiv 1$. In that case, bounds for the frequency $\omega$ can be
found. These read \cite{vw}
\begin{equation}
 \omega^2 \leq \omega^2_{\rm max} = \frac{1}{2} U''(0) = 1
\end{equation}
such that $V''(0) < 0$ and hence $\phi(0)=0$ corresponds to a maximum of the potential
and 
\begin{equation}
 \omega^2 \geq \omega^2_{\rm min} = 0  \ .
\end{equation}
This latter condition on the frequency comes from the requirement that the potential $V(\phi)$
is positive for some $\phi\neq 0$, i.e. $\omega^2 \phi^2 > 1-\exp(-\phi^2)$ for some $\phi\neq 0$. 
In contrast to the $\phi^6$-potential used in \cite{vw,kk1} we can always find a $ \phi\neq 0$ for any choice
of $\omega^2 \in [0:1]$ that fulfills this condition.

\begin{figure}[h]
\begin{center}

\subfigure[][potential $V$ for $\Lambda=-0.1$, $r=10$]{\label{pot1}
\includegraphics[width=5.5cm,angle=270]{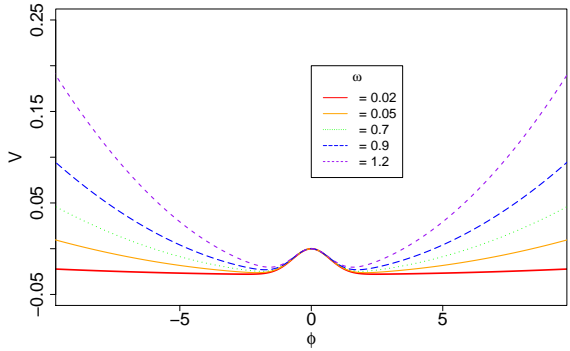}}
\subfigure[][potential $V$ for $\omega=0.3$, $r=10$]{\label{pot2}
\includegraphics[width=5.5cm,angle=270]{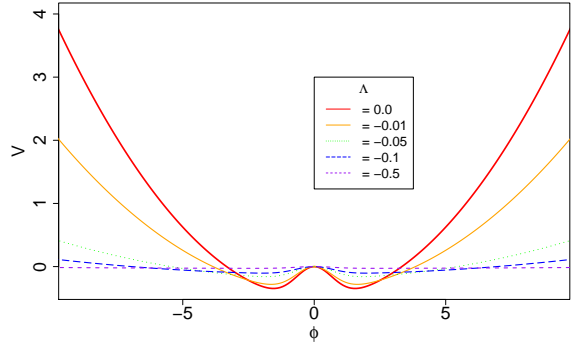}} \\
\subfigure[][potential $V$ for $\omega=0.3$ and $\Lambda=-0.1$]{\label{pot3}
\includegraphics[width=5.5cm,angle=270]{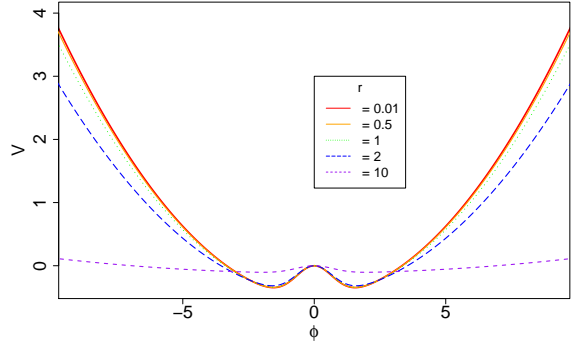}}
\end{center}
\caption{\label{potentialV} We show the effective potential $V(\phi)$ for $Q$-balls
in an AdS background for fixed $r=10$, $\Lambda=-0.1$ and different values of $\omega$ (a),
for fixed $r=10$, $\omega=0.3$ and different values of $\Lambda$ (b), for $\omega=0.3$, $\Lambda=-0.1$ and
different values of the radial coordinate $r$ (c).}
\end{figure}

As soon as either $\Lambda$ or $G$ are non-vanishing the metric function $N$ depends on $r$. In Fig.\ref{potentialV}
we show the potential $V(\phi)$ for $G=0$, different values of $\Lambda$, $\omega$ 
and of the ``time''-coordinate $r$. Fig.\ref{pot1} and Fig.\ref{pot2} show the potential for fixed $r=10$.
In Fig.\ref{pot1} the cosmological constant is fixed $\Lambda=-0.1$ and $\omega$ varied. We observe that
the potential becomes more narrow when increasing $\omega$. In Fig.\ref{pot2} we show the potential
for $\omega=0.3$ and different values of $\Lambda$. Here, we observe that the potential flattens when decreasing
$\Lambda$. Finally, in Fig.\ref{pot3} we show the potential for $\omega=0.3$, $\Lambda=-0.1$ and different
values of the ``time''-coordinate $r$. As expected, the potential flattens for increasing $r$ such that
in the limit $r\rightarrow \infty$ the potential tends to zero.  

It is important to note that in contrast to the $\phi^6$-potentials used in previous studies of $Q$-balls in
Minkowski space-time \cite{vw} our potential does not possess local maxima at some $\phi\neq 0$. 

We have to construct the solutions to (\ref{eqnew}) numerically. We have done this applying a 
Newton-Raphson method that uses an adaptive grid scheme \cite{colsys}.

\subsection{Numerical results}
We have first studied the dependence of the mass $M$ and charge $Q$ on $\omega$ for different
values of $\Lambda$. This is shown in Fig.\ref{qballs}. We observe that the mass $M$ and charge $Q$ 
of the $Q$-balls diverge at $\omega^2_{\rm min}$, respectively
$\omega^2_{\rm max}$ in asymptotically flat space-time, i.e. for $\Lambda=0$. This is very similar to the
qualitative behaviour in the case of a $\phi^6$-potential \cite{vw,kk1,kk2}. 
Decreasing the cosmological constant and hence considering $Q$-balls in an AdS background
space-time that does not backreact, we find that $M$ and $Q$ still diverge at $\omega^2_{\rm min}$.
Moreover, we find that $\omega^2_{\rm min}$ does not seem to depend on $\Lambda$.
However, at $\omega^2_{\rm max}$ we find that now both $Q$ and $M$ tend to zero - very similar to what happens
for self-gravitating $Q$-balls, so-called boson stars (see e.g. \cite{kk1,kk2}). 
Moreover, $\omega^2_{\rm max}$ increases with decreasing $\Lambda$, e.g. we find
$\omega^2_{\rm max}=1$ for $\Lambda=0$, $\omega^2_{\rm max}\approx 1.16$ for $\Lambda=-0.03$ and
$\omega^2_{\rm max}\approx 1.21$ for $\Lambda=-0.05$.
At $\omega^2_{\rm max}$ the value of $\phi(0)=0$ and hence $\phi(r)\equiv 0$. This is shown
in Fig. \ref{qballs2} and is, of course,
related to the fact that for $\omega^2 > \omega^2_{\rm max}$ the potential $V(\phi)$ no longer
possesses a maximum at $\phi=0$, but rather a minimum. In flat space-time ($\Lambda=0$) the profile
of $\phi$ would hence become more and more spread out over all space and just before becoming
zero everywhere would be nearly constant (and non-vanishing) on a larger domain of space-time.
This is what makes the integrals for $Q$ and $M$ diverge. On the other hand, as soon as $\Lambda\neq 0$ the AdS
background acts as a confining box and the integral never diverges. 
At $\omega^2_{\rm min}$ we find that $\phi(0)$ tends to very large values. This is
see in Fig.\ref{qballs2}. The profile
of $\phi(r)$ becomes hence very sharply peaked on a small interval of $r$ which makes the
integrals for the charge and mass diverge, even if $\Lambda\neq 0$. 

This changed behaviour of the charge and mass at $\omega^2_{\rm max}$ for $\Lambda\neq 0$ 
now leads to qualitatively
different plots for the charge $Q$ over the mass $M$. This is shown in Fig.\ref{qballs3}, where
we show $Q$ as function of $M$ and compare $M$ with the mass of $Q$ free boson $m_{\rm B}=1$ (in our rescaled
units). 
While for $\Lambda=0$ there exist two solutions with different $M$ for a given charge, one of which
has $M < m_{\rm B} Q$ (stable branch) and the other $M > m_{\rm B} Q$ (unstable branch), there exists only one branch of solutions as soon as $\Lambda\neq 0$. The $Q$-balls in AdS space-time
exist for arbitrarily small values of the mass and the charge. We observe that the mass drops below $M=m_{\rm B} Q$ for sufficiently
large values of the charge. We would hence expect that the $Q$-balls are unstable to a decay into
$Q$ free bosons with mass $m_{\rm B}$ for small charges, but stable against this decay for large charges $Q$.
We also observe that the critical value of $Q$, $Q_{\rm crit}$ at which the mass of the $Q$-ball is
$M=m_{\rm B} Q_{\rm crit}$ depends on $\Lambda$. The smaller $\Lambda$ the larger $Q_{\rm crit}$ such that
for $Q < Q_{\rm crit}$ the $Q$-balls are unstable, while for $Q > Q_{\rm crit}$ the $Q$-balls are stable.
A smaller $\Lambda$ corresponds to a larger absolute value of $\Lambda$ and hence to a smaller AdS radius.

\begin{figure}[h]
\begin{center}

\subfigure[][charge $Q$ over $\omega$]{\label{charge_qball}
\includegraphics[width=5.5cm,angle=270]{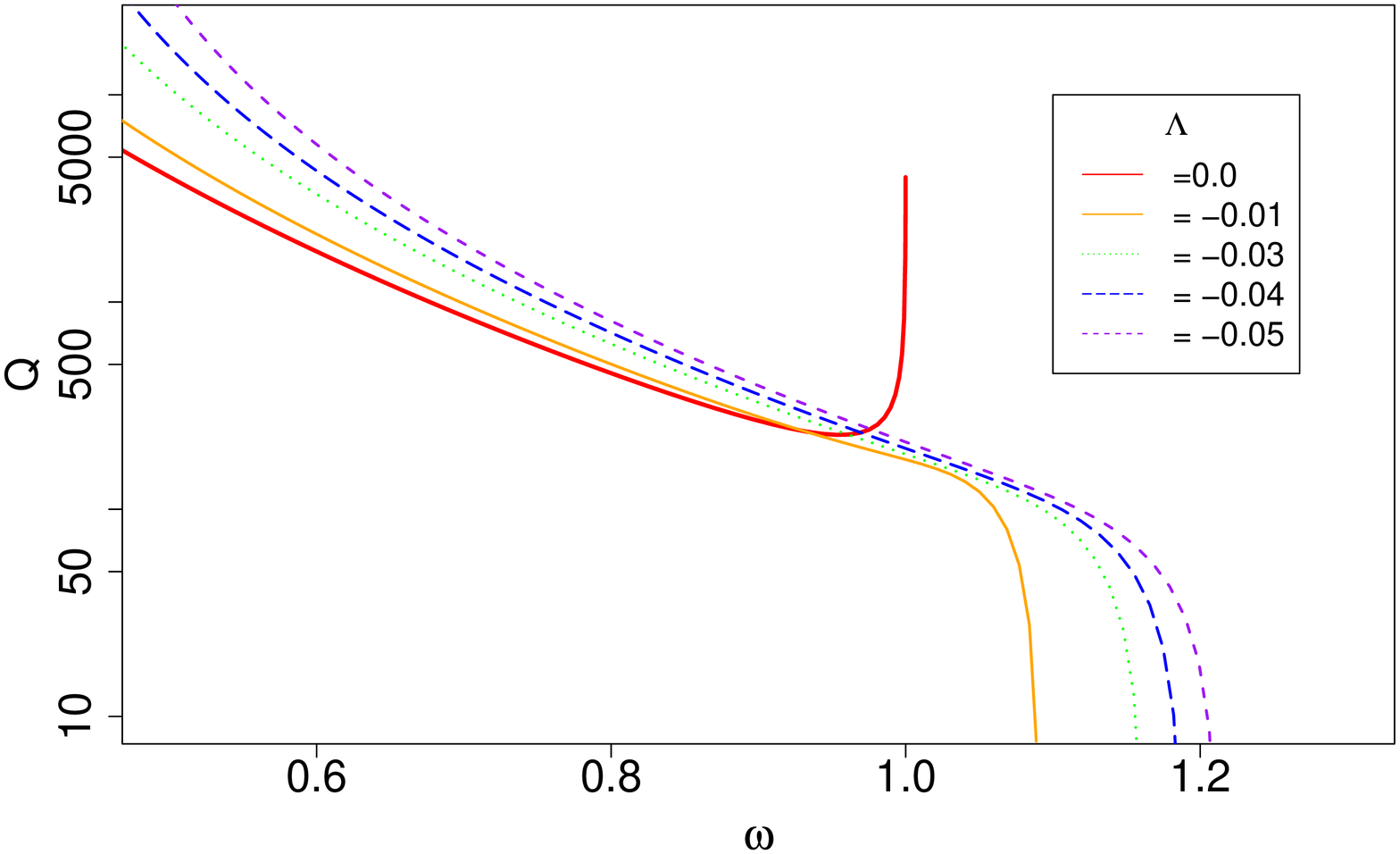}}
\subfigure[][mass $M$ over $\omega$]{\label{mass_qball}
\includegraphics[width=5.5cm,angle=270]{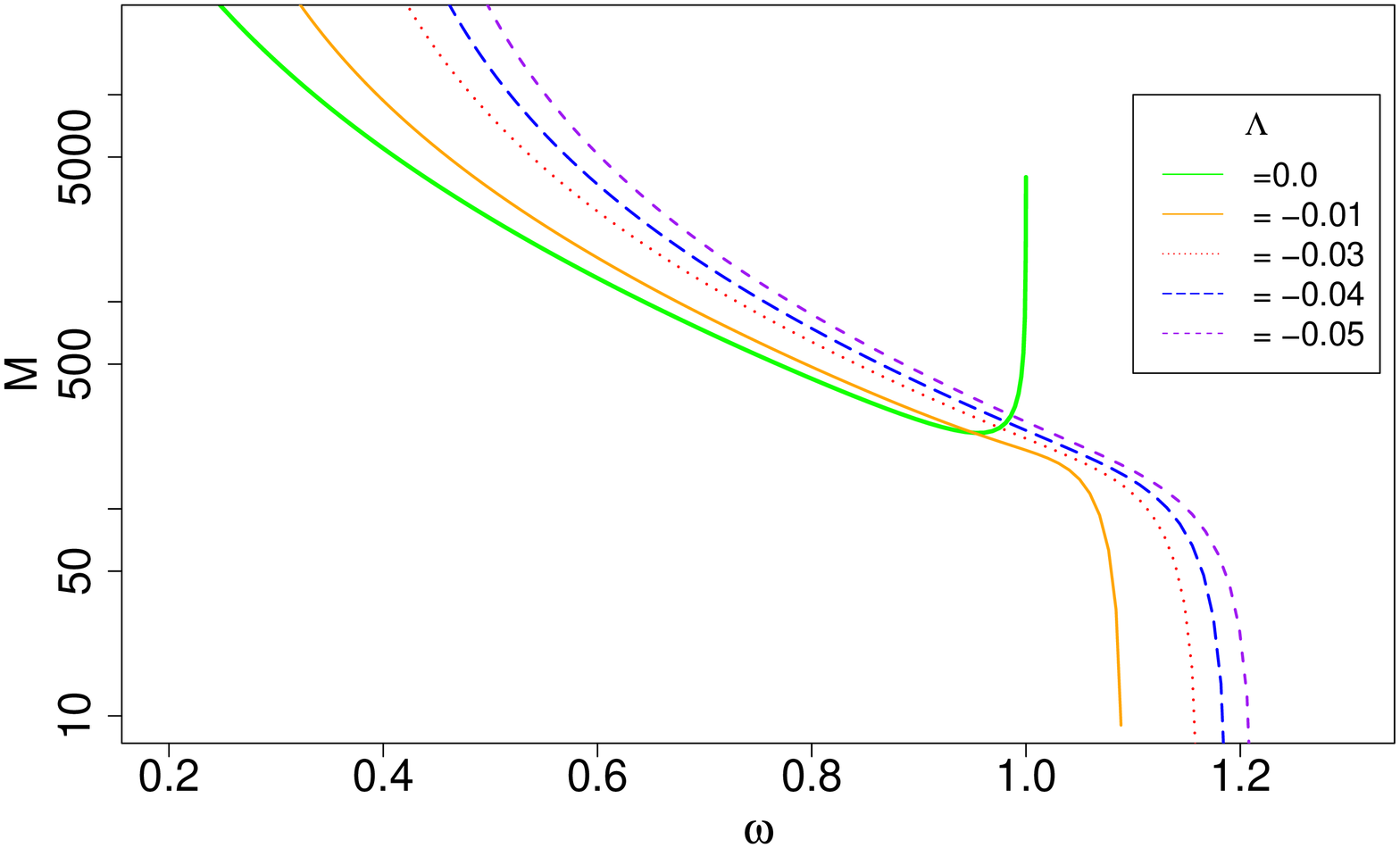}} 
\end{center}
\caption{\label{qballs} We show the charge $Q$ (left) as well as the mass $M$ (right)
of $Q$-balls in an Anti-de Sitter background for different values of the
cosmological constant $\Lambda$ in dependence on the frequency $\omega$. For comparison we also show the values for
$Q$-balls in a flat space-time background ($\Lambda=0$). }
\end{figure}

\begin{figure}[h]
\begin{center}

\subfigure[][charge $Q$ over $\phi(0)$]{\label{charge_qball2}
\includegraphics[width=5.5cm,angle=270]{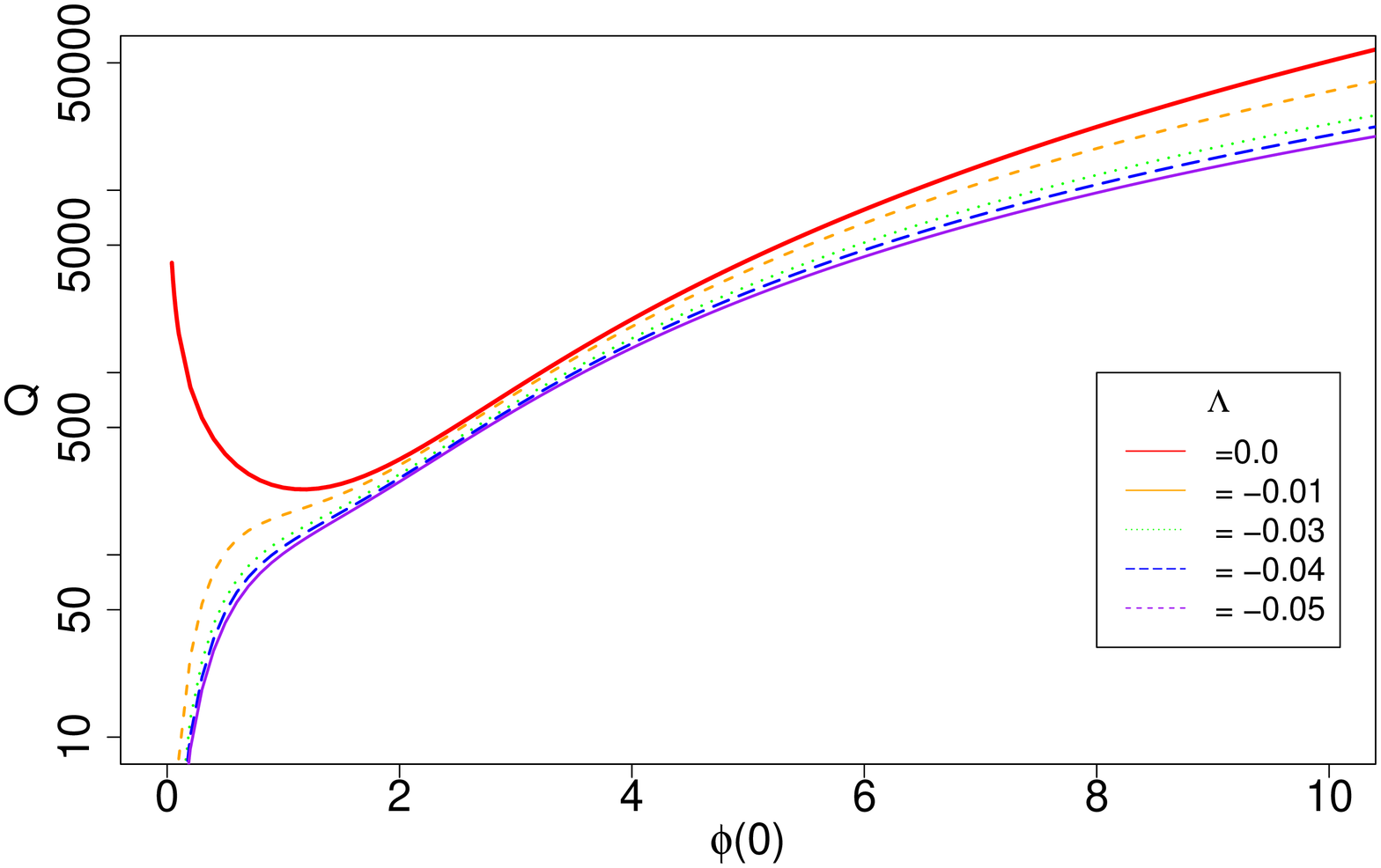}}
\subfigure[][mass $M$ over $\phi(0)$]{\label{mass_qball2}
\includegraphics[width=5.5cm,angle=270]{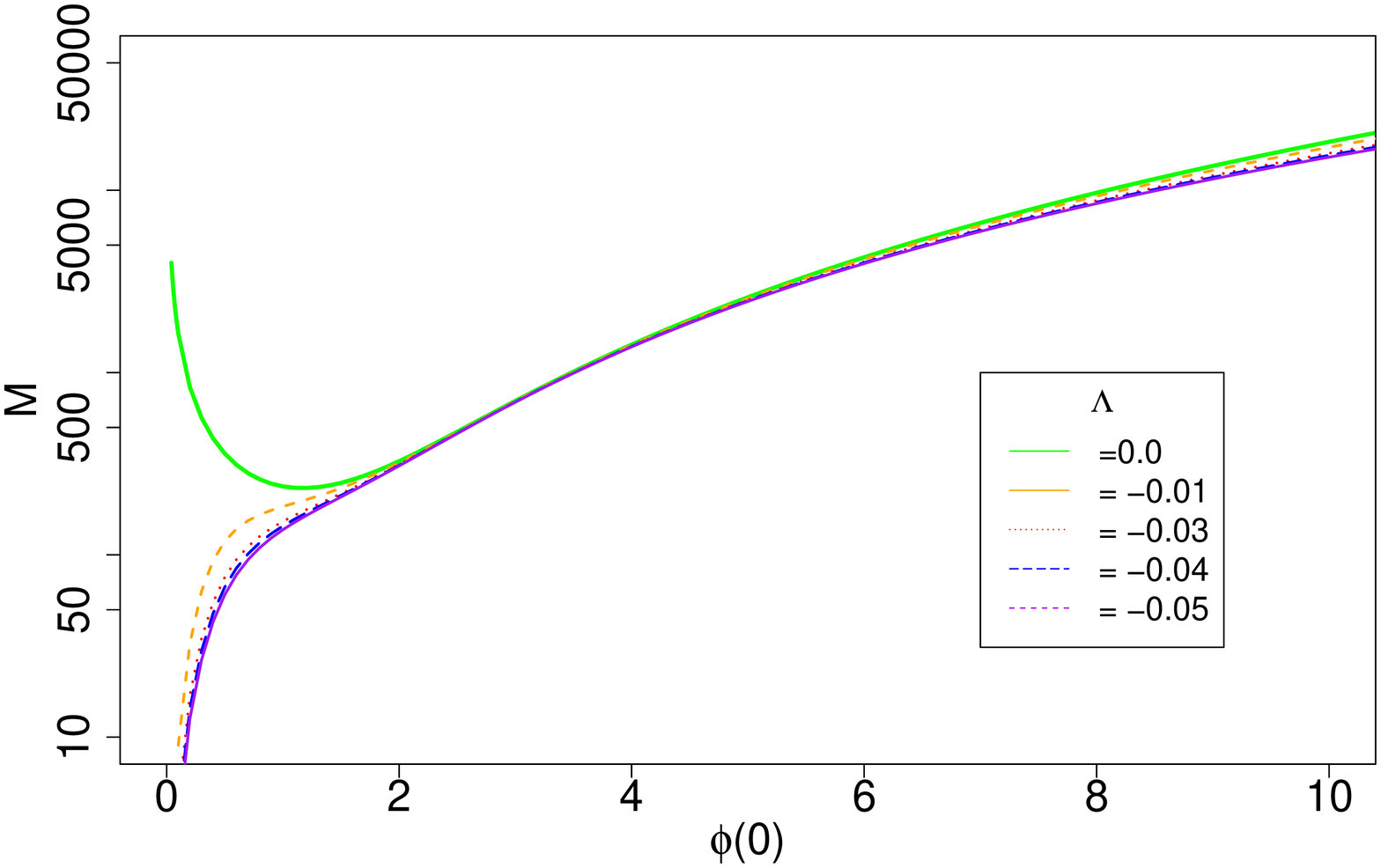}} 
\end{center}
\caption{\label{qballs2} We show the charge $Q$ (left) as well as the mass $M$ (right)
of $Q$-balls in an Anti-de Sitter background for different values of the
cosmological constant $\Lambda$ in dependence on the value of the scalar function
$\phi$ at the origin, $\phi(0)$. For comparison we also show the values for
$Q$-balls in a flat space-time background ($\Lambda=0$). }
\end{figure}

\begin{figure}[h]
\begin{center}
\includegraphics[width=7cm,angle=270]{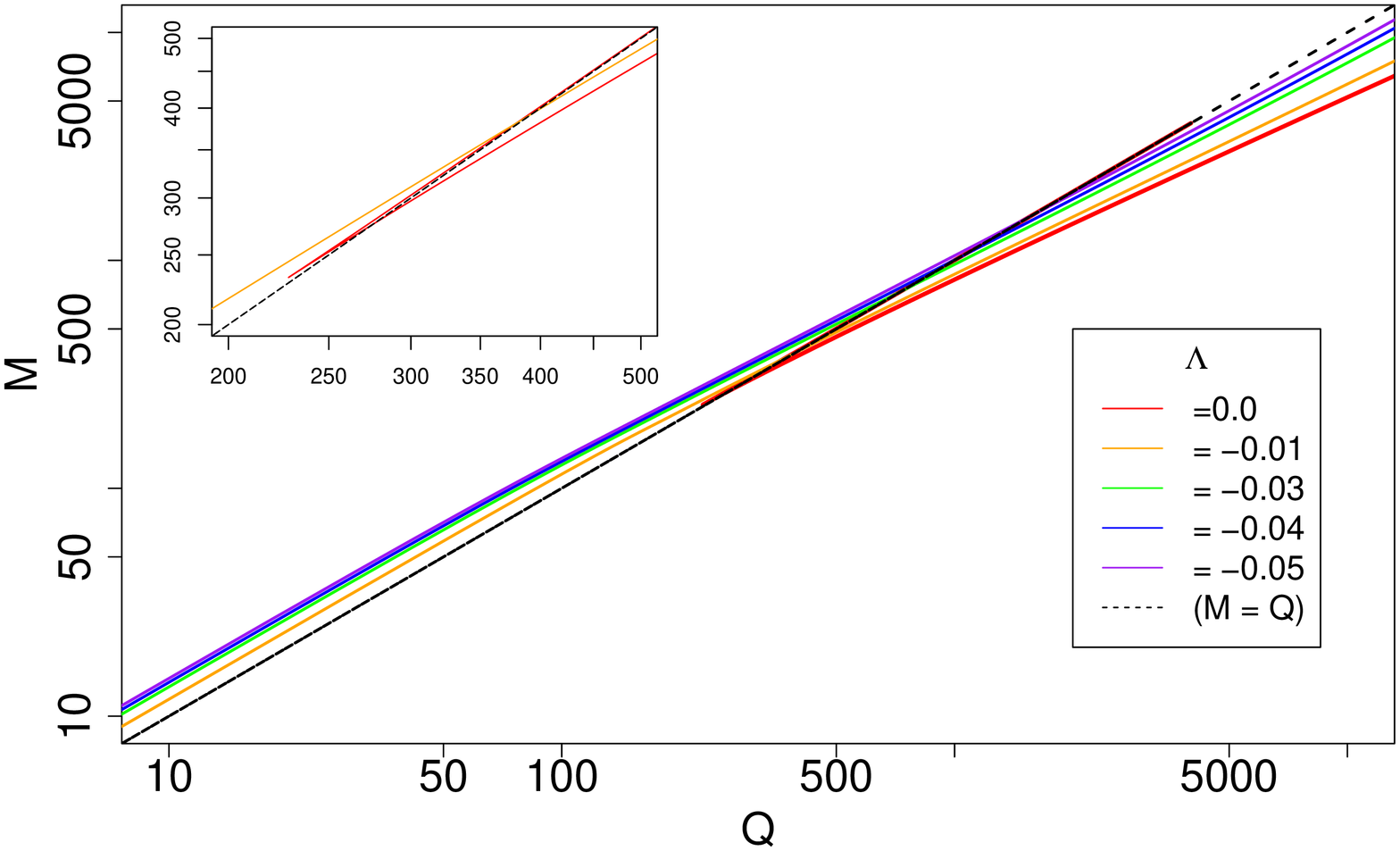}
\end{center}
\caption{\label{qballs3} We show the mass $M$ as function of the
charge $Q$ for $Q$-balls in an Anti-de Sitter background for different values of the
cosmological constant $\Lambda$. For comparison we also show the values for
$Q$-balls in a flat space-time background ($\Lambda=0$). Note that the line $M=Q$ corresponds
to the mass of $Q$ free boson $M=m_{\rm b} Q$ with mass $m_{\rm b}=1$ (in our rescaled variables).}
\end{figure}

Next to the fundamental solutions presented above, it has been observed previously \cite{kk1,kk2} that $Q$-ball solutions
with a finite number $k$, $k\in \mathbb{N}$, of nodes of the scalar field function $\phi(r)$ exist. These solutions are typically
interpreted as radially excited solutions. We also find these solutions for our scalar potential
and when including a negative cosmological constant. 
In Fig.\ref{nodes_profiles}
we show the profiles of the solutions with $k=0,1,2$ nodes for $\Lambda=-0.01$, where the $k=0$ solution
corresponds to the fundamental $Q$-ball solution discussed above. We observe that the smallest value of $r$ at which
$\phi(r)$ becomes zero decreases with increasing $k$. Moreover, the zeros appear at small values of $r$, i.e.
close to the core of the $Q$-ball.

\begin{figure}[h]
\begin{center}
\includegraphics[width=7cm,angle=270]{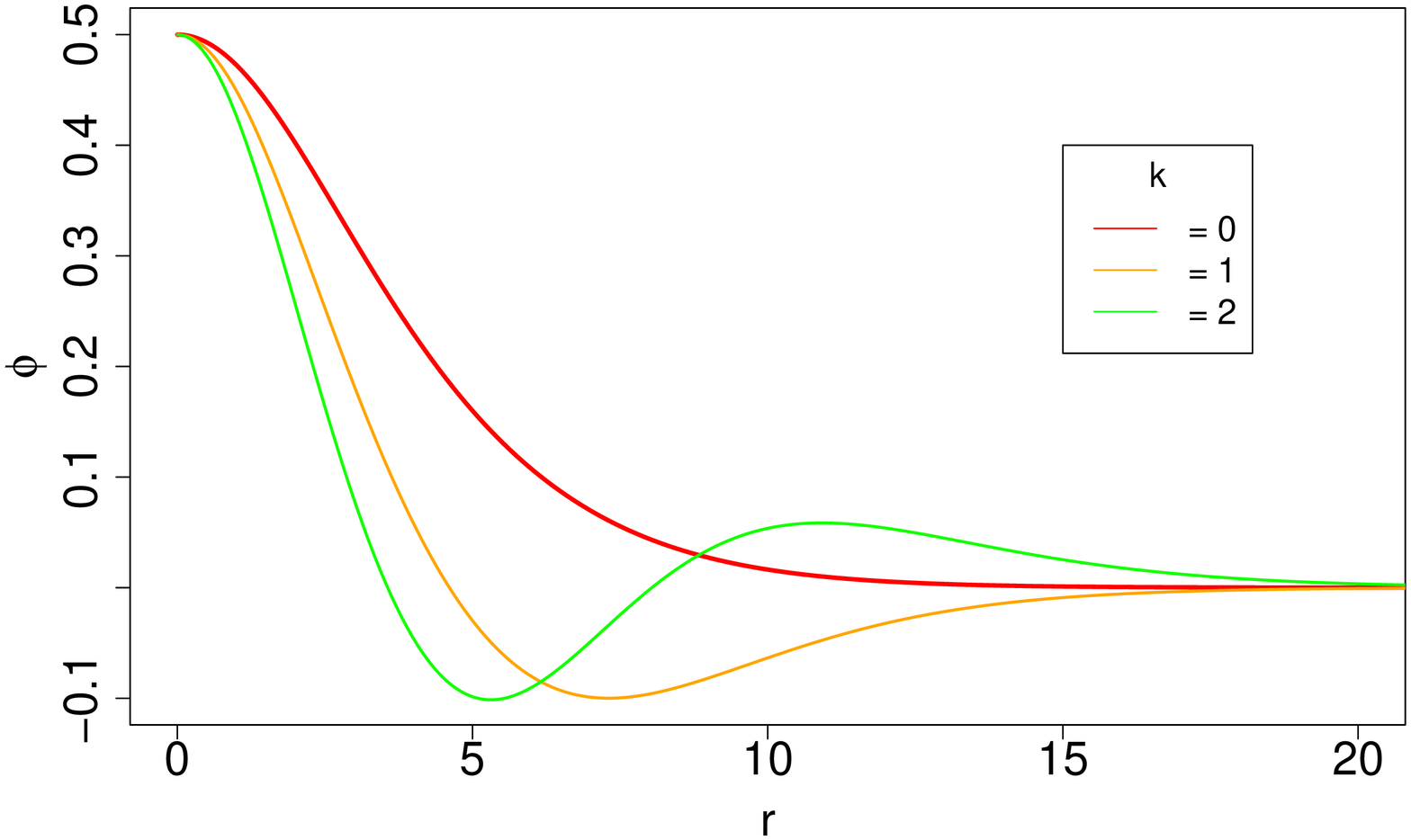}
\end{center}
\caption{\label{nodes_profiles}  We show the profile of the scalar field function $\phi(r)$
for $Q$-balls with $k=0,1,2$ nodes, respectively, for $\Lambda=-0.01$.  }
\end{figure}

\begin{figure}[h]
\begin{center}

\subfigure[][charge $Q$ over $\omega$]{\label{charge_qballEX}
\includegraphics[width=5.5cm,angle=270]{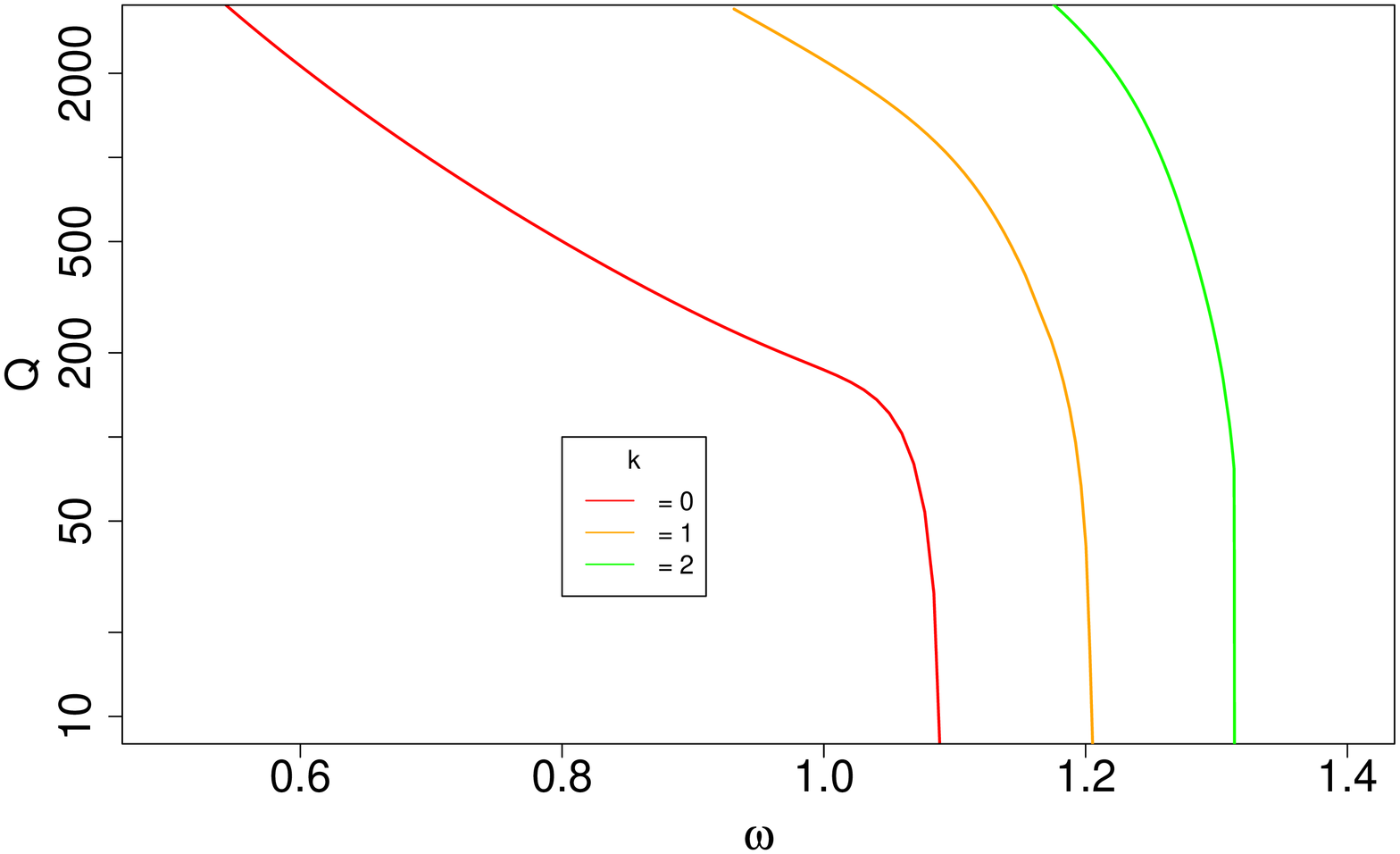}}
\subfigure[][mass $M$ over $\omega$]{\label{mass_qballEX}
\includegraphics[width=5.5cm,angle=270]{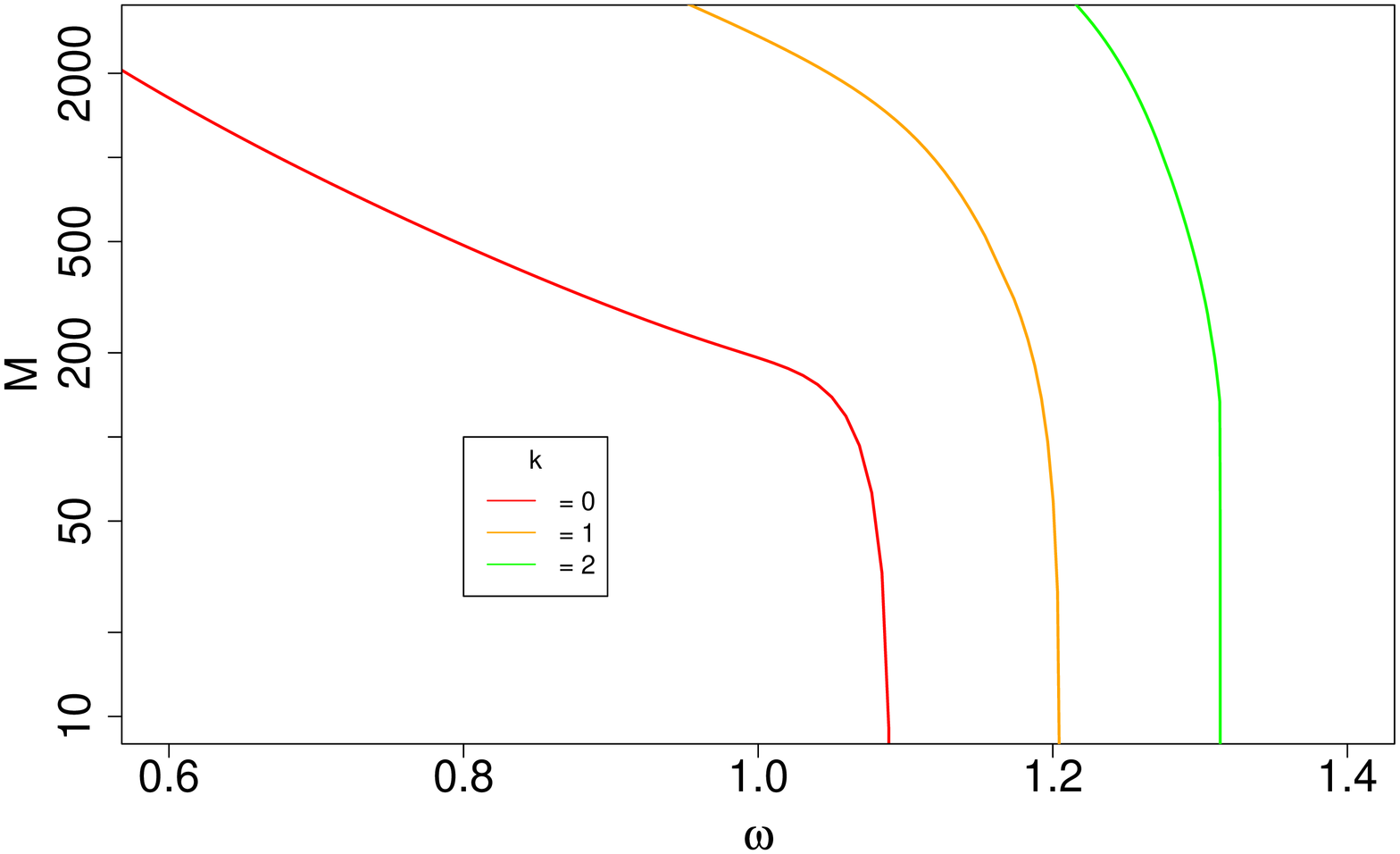}} 
\end{center}
\caption{\label{qballsEX} We show the charge $Q$ (left) as well as the mass $M$ (right)
of $Q$-balls in an Anti-de Sitter background for $\Lambda=-0.01$ in dependence 
on the frequency $\omega$ for the $k=0,1,2$ solutions. }
\end{figure}

\begin{figure}[h]
\begin{center}
\includegraphics[width=7cm,angle=270]{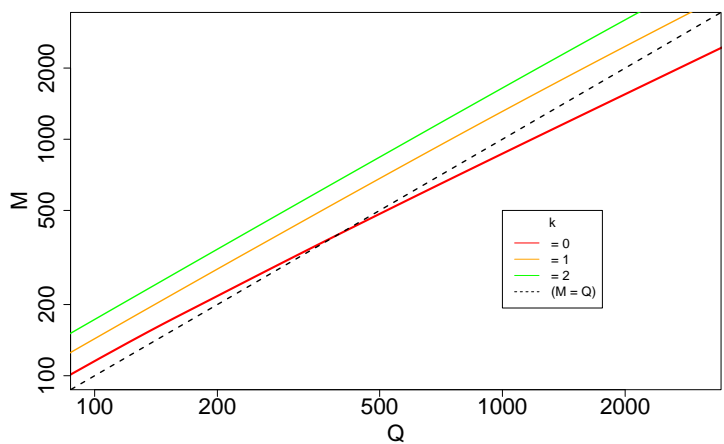}
\end{center}
\caption{\label{qballsEX2} We show the mass $M$ as function of the
charge $Q$ for $Q$-balls in an Anti-de Sitter background with $\Lambda=-0.01$ for the $k=0,1,2$ solutions.
The dashed line is $M=Q$ and corresponds to the mass of $Q$ free bosons with mass $m_{\rm B}=1$ (in our
rescaled variables). }
\end{figure}

\begin{figure}[h]
\begin{center}

\subfigure[][$\kappa=0.005$: charge $Q$ over $\omega$]{\label{charge_BS1}
\includegraphics[width=5.5cm,angle=270]{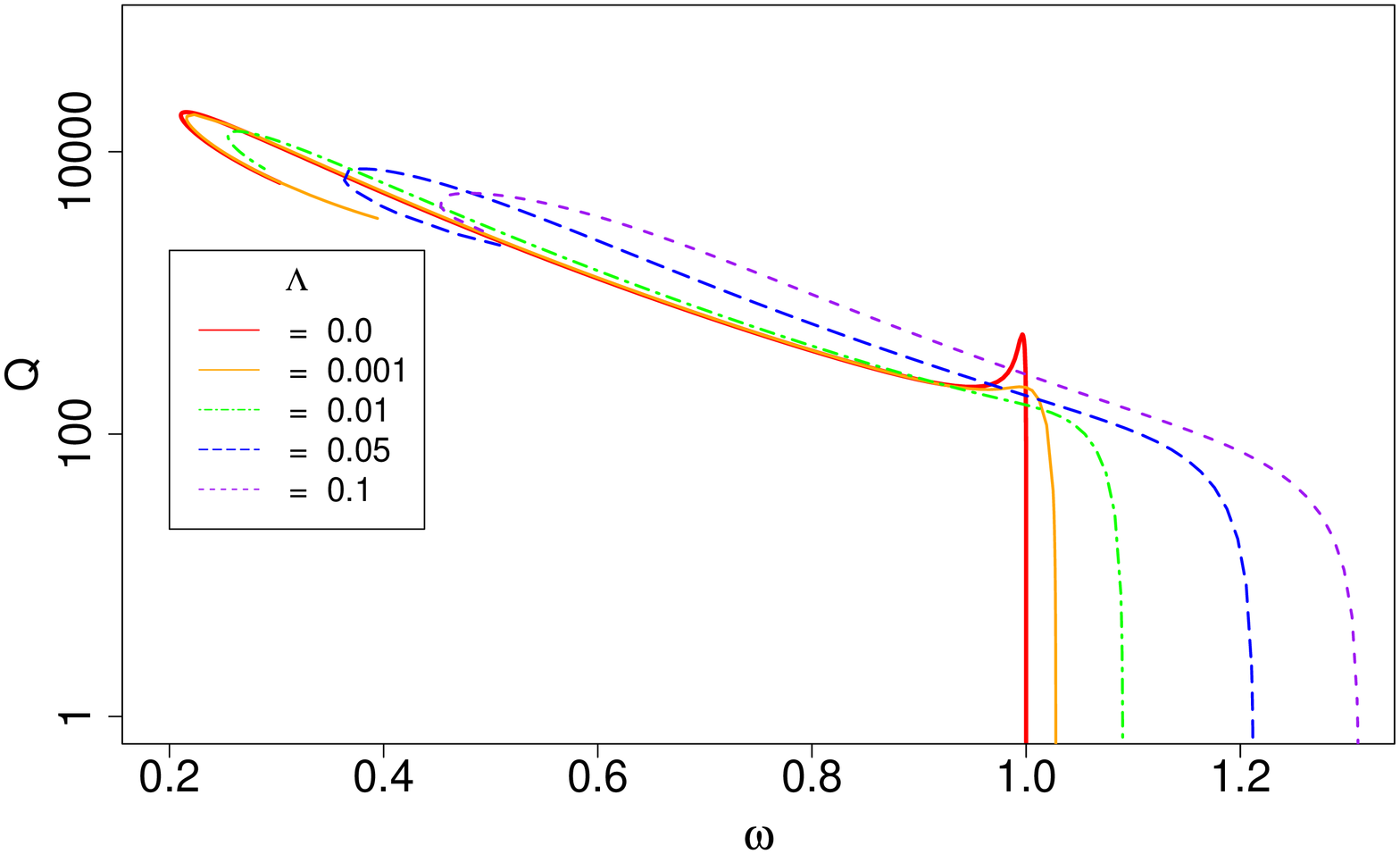}}
\subfigure[][$\kappa=0.005$: mass $M$ over $\omega$]{\label{mass_BS1}
\includegraphics[width=5.5cm,angle=270]{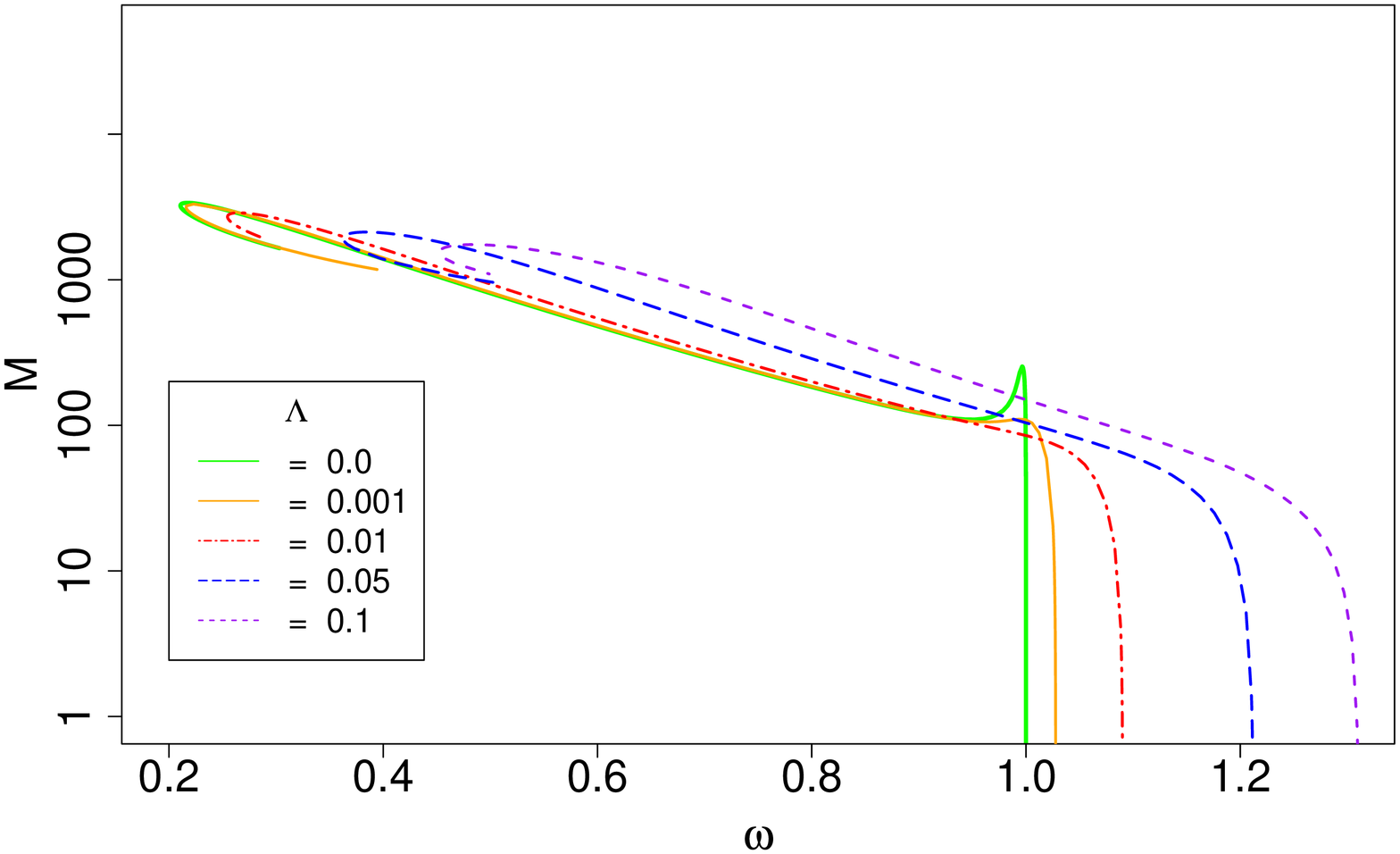}} \\
\subfigure[][$\kappa=0.01$: charge $Q$ over $\omega$]{\label{charge_BS2}
\includegraphics[width=5.5cm,angle=270]{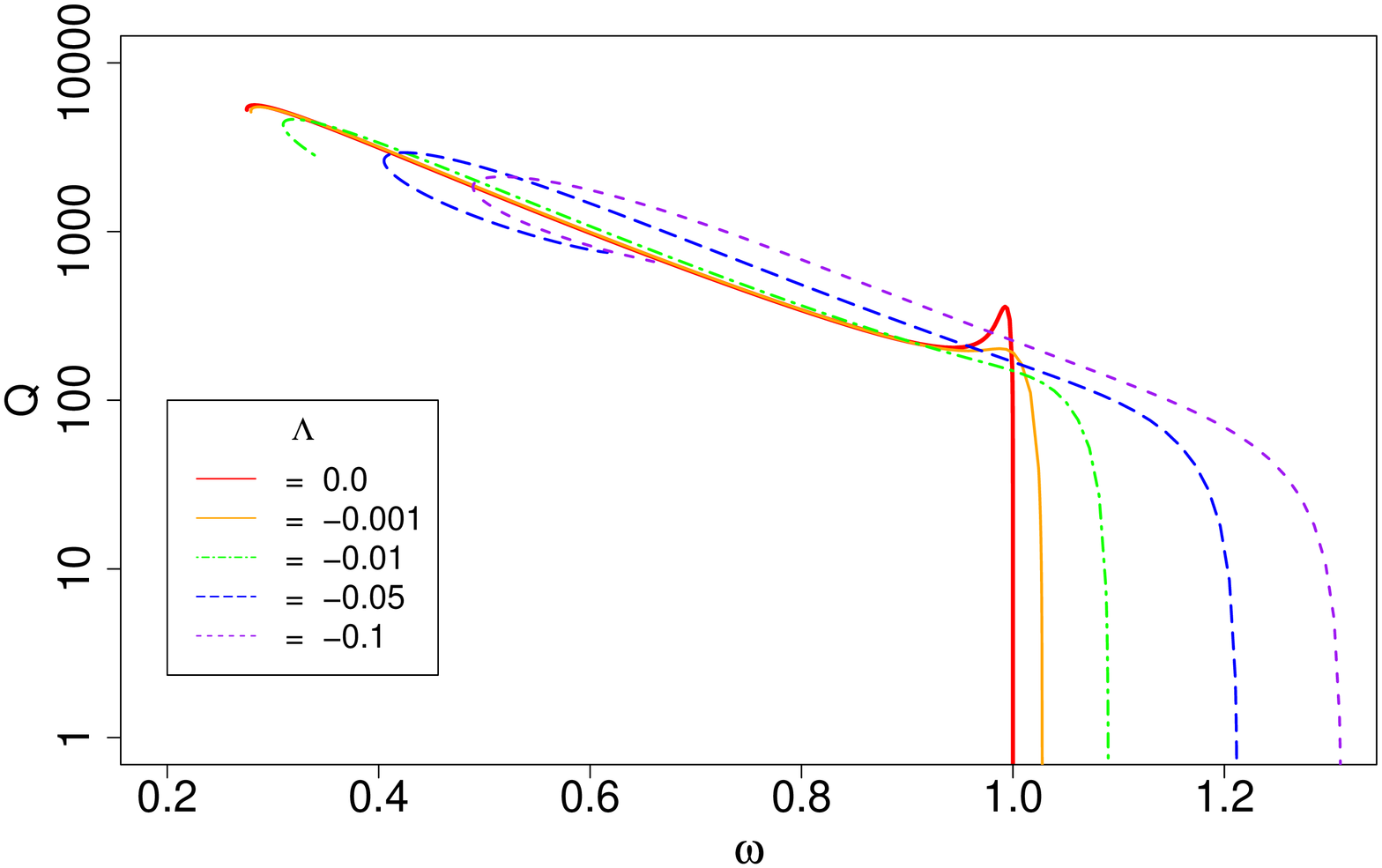}}
\subfigure[][$\kappa=0.01$: mass $M$ over $\omega$]{\label{mass_BS2}
\includegraphics[width=5.5cm,angle=270]{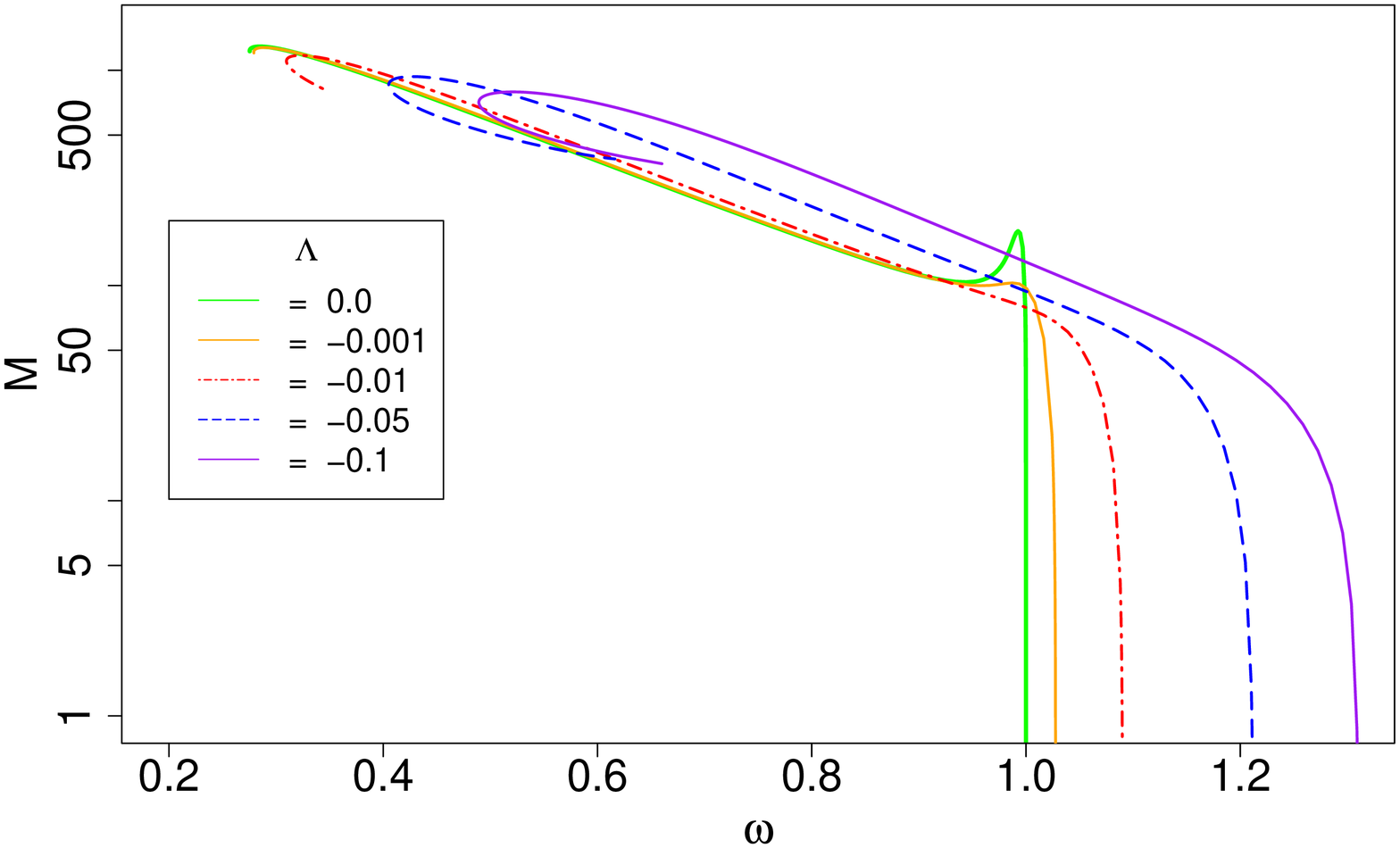}} 

\end{center}
\caption{\label{BS1} We show the charge $Q$ (left) as well as the mass $M$ (right)
of boson stars in Anti-de Sitter space-time for different values of the
cosmological constant $\Lambda$ in dependence on the frequency $\omega$ for $\kappa=0.005$ (top)
and $\kappa=0.01$ (bottom), respectively.  }
\end{figure}

Now it is interesting to see how the mass $M$ and charge $Q$ of the $Q$-ball solutions with nodes
depend on $\omega$. This is shown in Fig.\ref{qballsEX} for $\Lambda=-0.01$ and $k=0,1,2$, respectively.
Our results suggest that the maximal value of $\omega$ up to which the solutions exist
depends on the number of nodes. We find that $\omega_{\rm max}$ increases with increasing node number, i.e.
radially excited $Q$-ball solutions exist for larger values of the frequency $\omega$. 
Moreover, both the charge (see Fig.\ref{charge_qballEX}) as well as the mass (see Fig.\ref{mass_qballEX})
increase with $k$ for a fixed value of $\omega$. This is not surprising since these are radially excited
solutions. 
In Fig.\ref{qballsEX2} we show the mass as function of the charge for $Q$-balls in an AdS background with
$\Lambda=-0.01$ and for $k=0,1,2$. Clearly, for a given 
charge $Q$ the mass $M$ of the solutions increases with increasing $k$. Moreover, the mass $M$ of the $k=1,2$ solutions
is always above that of $Q$ free bosons with mass $m_{\rm B}=1$, while for $k=0$ it is above that curve
for small $Q$ and below for large $Q$. Of course, we would expect the excited solutions to be unstable and
our numerical results here confirm exactly this expectation.

\section{Boson stars in AdS space-time}
Here, we choose $\kappa\neq 0$. In this case, we have to solve the full set of
coupled and non-linear equations (\ref{eq1})-(\ref{eq3}) numerically. Again, we have done this
with a Newton-Raphson method \cite{colsys}.

\subsection{Numerical results}
In Fig.\ref{BS1} we show the charge $Q$ and $M$ of the boson stars in dependence on the frequency $\omega$ 
for different values of $\Lambda$ and two different values of $\kappa$.

We observe that the local maximum of the charge and mass, respectively, that exists close to $\omega_{max}$ for $\Lambda=0$ and
$\kappa\neq 0$ disappears as soon as $\Lambda\neq 0$. Moreover, as for $\kappa=0$ the value of $\omega_{max}$ at which
the charge and mass become zero does not depend on $\kappa$. This has been observed before in asymptotically
flat space-times (see e.g. \cite{kk1,kk2}) and seems to be true also for $\Lambda\neq 0$.
For $\kappa\neq 0$ the solutions exist down to a minimal
finite value of $\omega=\omega_{\rm min} > 0$. This $\omega_{\rm min}$ depends on $\kappa$ as well as on $\Lambda$.
It increases with increasing $\kappa$ and increases with decreasing $\Lambda$, i.e. the larger the curvature of the space-time
and the larger the interaction between the curvature and the matter content, the larger $\omega_{\rm min}$. 
At $\omega_{\rm min}$ a second branch of solutions exists that has smaller mass and charge than the first (main) branch of solutions.
We have only constructed the second branch of solutions here, but believe that in analogy to boson stars
in asymptotically flat space-time several further branches exist that form a spiral.

\begin{figure}[h]
\begin{center}

\subfigure[][$\Lambda=0$]{\label{charge_BS4}
\includegraphics[width=5.5cm,angle=270]{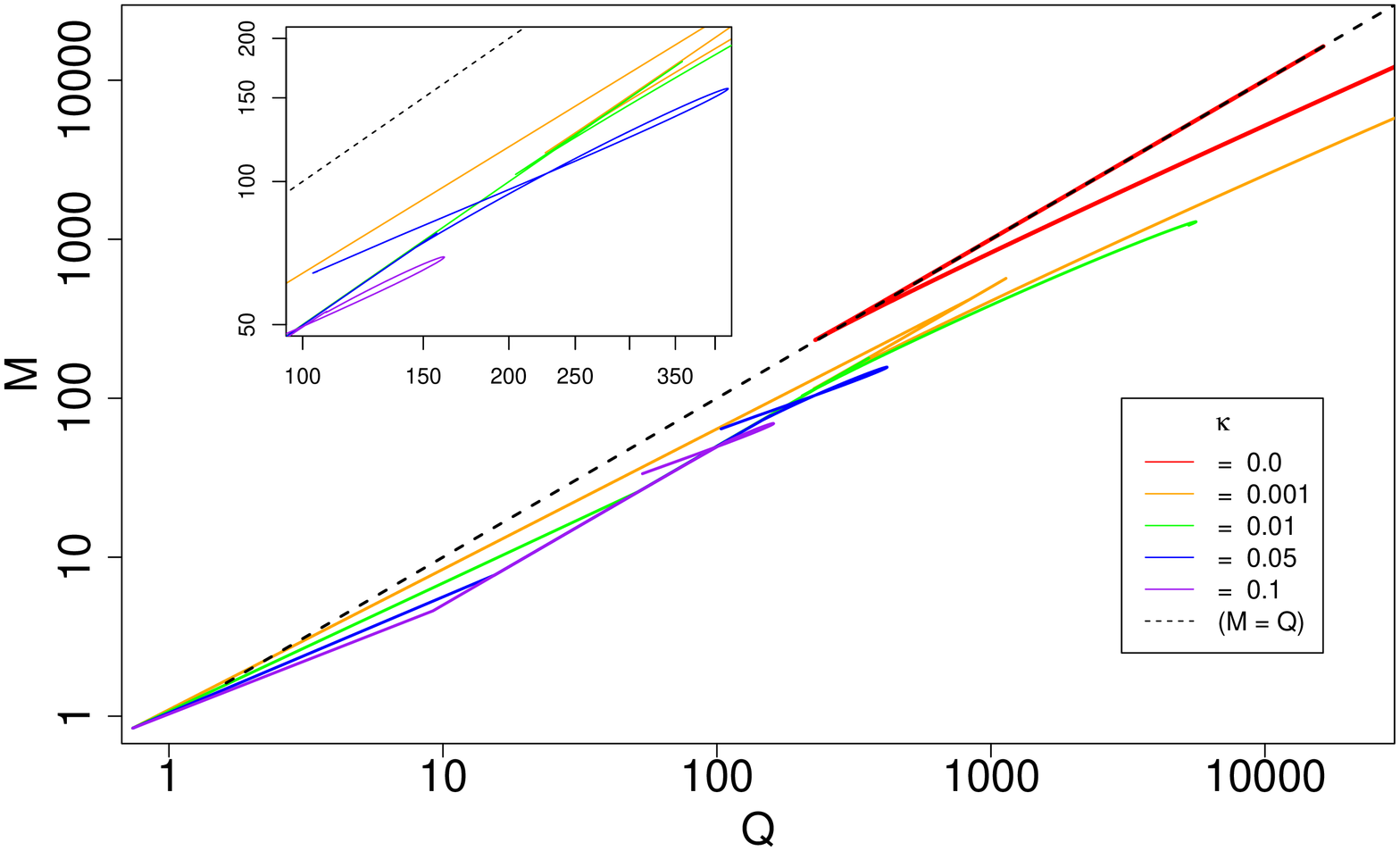}}
\subfigure[][$\Lambda=-0.001$]{\label{mass_BS5}
\includegraphics[width=5.5cm,angle=270]{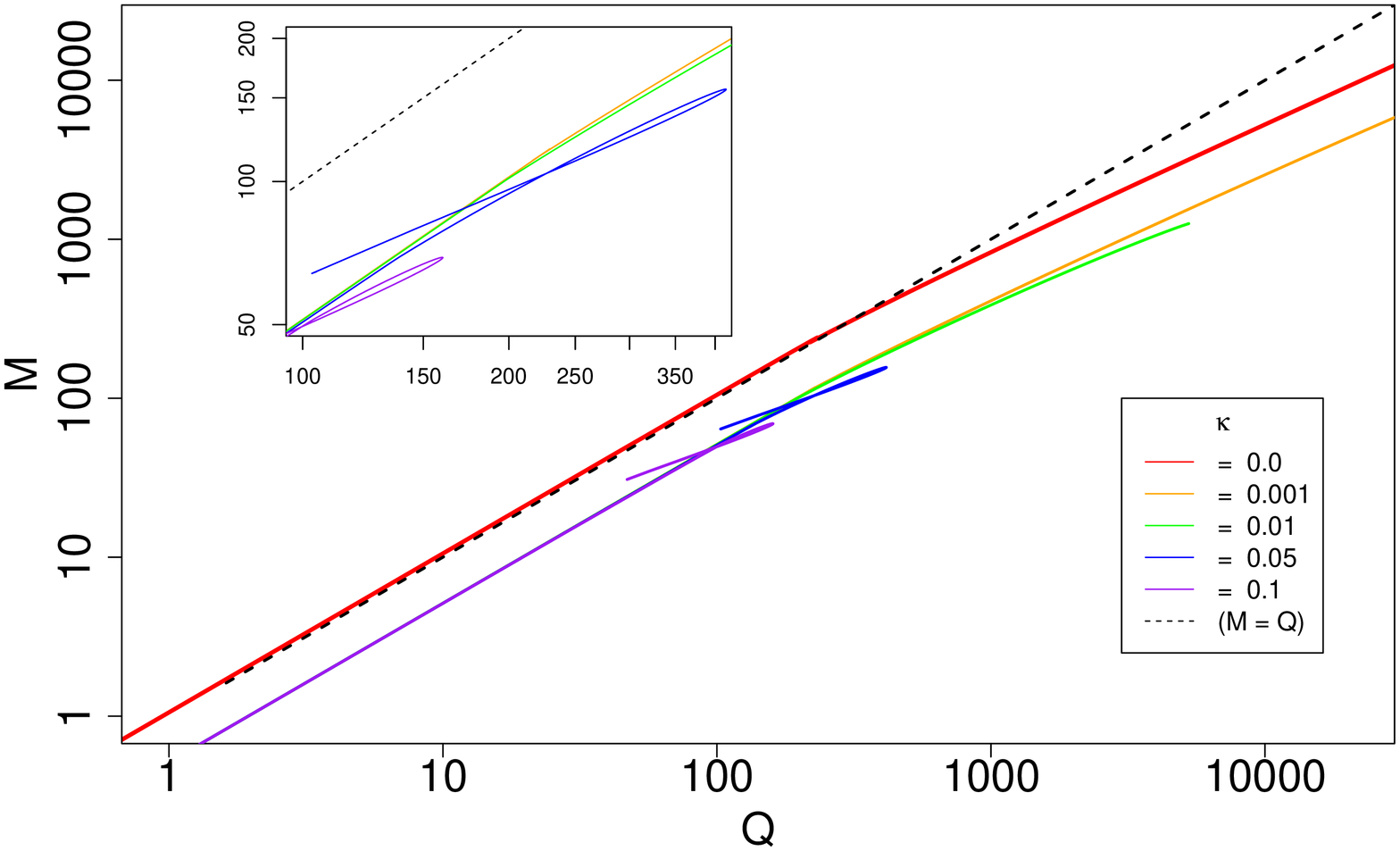}} \\
\subfigure[][$\Lambda=-0.01$]{\label{charge_BS6}
\includegraphics[width=5.5cm,angle=270]{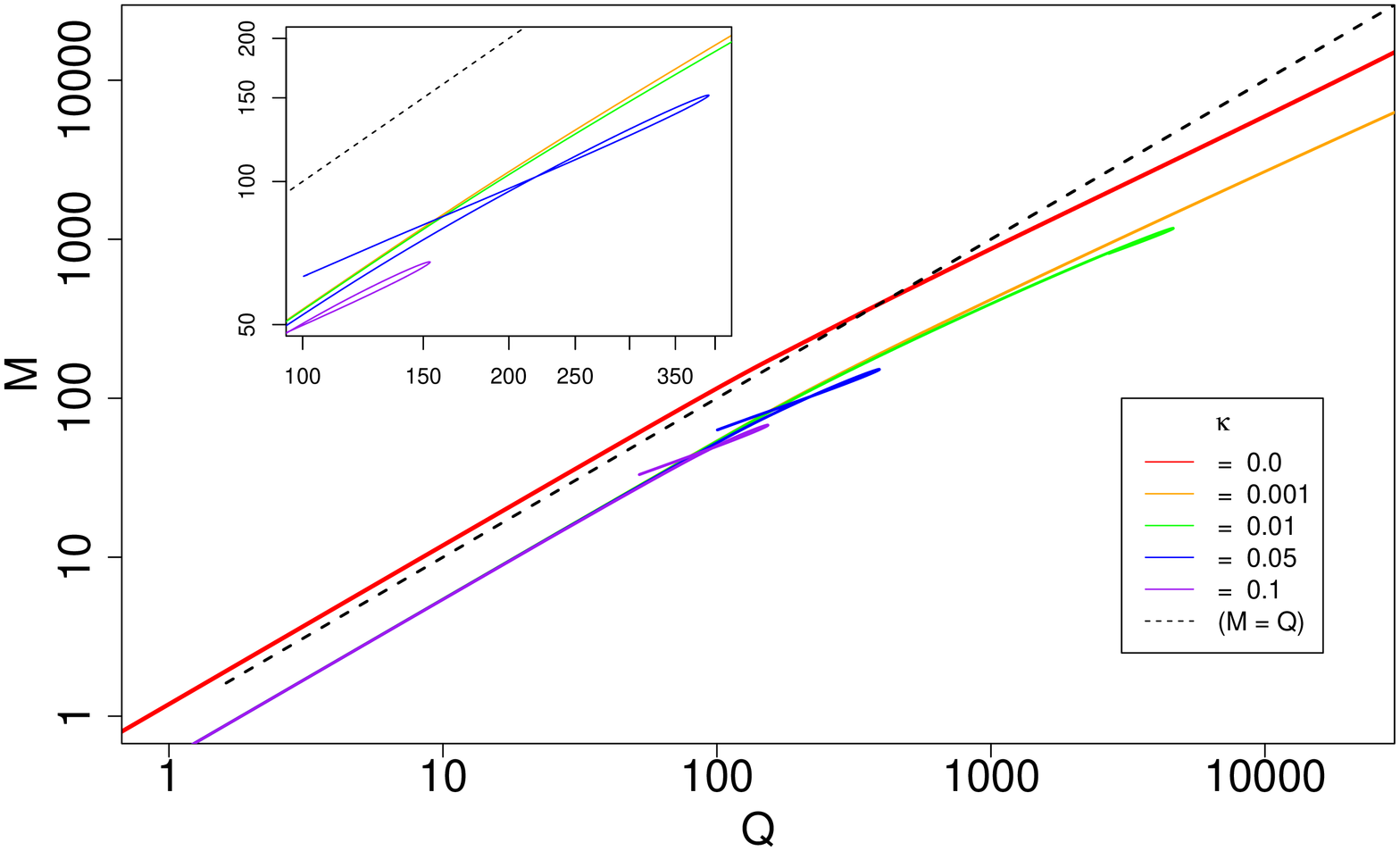}}

\end{center}
\caption{\label{BS2} We show the mass $M$ as function of the charge $Q$
of boson stars in Anti-de Sitter space-time for different values of $\kappa$ 
for $\Lambda=0$ (a), $\Lambda=-0.001$ (b) and $\Lambda=-0.01$ (c), respectively.}
\end{figure}

In Fig.\ref{BS2} we show the charge $Q$ as a function of the mass $M$ of boson stars
for different values of $\kappa$ and for the asymptotically flat case (see Fig.\ref{charge_BS4})
as well as for two asymptotically AdS cases with $\Lambda=-0.001$ (see Fig.\ref{mass_BS5}) and
$\Lambda=-0.01$ (see Fig.\ref{charge_BS6}), respectively. We observe that the features
of these plots change qualitatively when either choosing $\kappa$ or $\Lambda$ non-vanishing.
For $\kappa=\Lambda=0$, i.e. in the case of a fixed Minkowski space-time we find that
the solutions exist only down to a minimal, non-vanishing value of the charge and the mass and that
from there a second branch of solutions exists. The first branch of solutions has lower mass than
the mass of $Q$ free scalar boson with mass $m_{\rm B}$, while the second branch 
has (within our numerical accuracy) mass equal to the mass of $Q$ scalar bosons.
We would hence expect that the lower branch of $Q$-balls is stable, while the second branch possesses
an instability. Choosing $\kappa\neq 0$ for $\Lambda=0$, we find that the mass and charge of the boson
stars in asymptotically flat space-time are bounded from above, i.e. boson stars for the exponential
scalar potential employed here exist only up to a maximal value of the charge and mass. This is similar
to the observation for other potentials (see e.g. \cite{kk1,kk2}). Moreover, all solutions
have mass lower than that of $Q$ free scalar bosons. This is natural since gravity acts
to from a bound state out of the scalar bosons. Our results further suggest that the larger $\kappa$, the smaller
is the maximal value of the mass and charge, i.e. the closer the SUSY breaking scale is to the
Planck mass, the smaller the maximal mass and charge of the corresponding boson stars will be. 

For $\Lambda\neq 0$ and $\kappa=0$ the solutions exist for arbitrary values of the charge
and the mass and there is only one branch of solutions. For small $Q$ this branch of solutions is below
the $M=Q$ curve of $Q$ scalar bosons, while for large $Q$ it is below this curve. 
On the other hand, for $\Lambda\neq 0$ and $\kappa\neq 0$,
the solutions exist up to a maximal value of the charge and mass and from these values on, a second branch of
solutions exists. Again, the maximal value of the mass and charge depend strongly on $\kappa$, while
the dependence on $\Lambda$ seems to be very small (at least for the values of $\Lambda$ that we have chosen here).
Interestingly, the solutions on both branches have mass below $M=Q$.

\begin{figure}[h]
\begin{center}

\subfigure[][$<O>^{\frac{1}{\Delta}}$ over $\phi(0)$]{\label{holo1_1}
\includegraphics[width=6cm,angle=270]{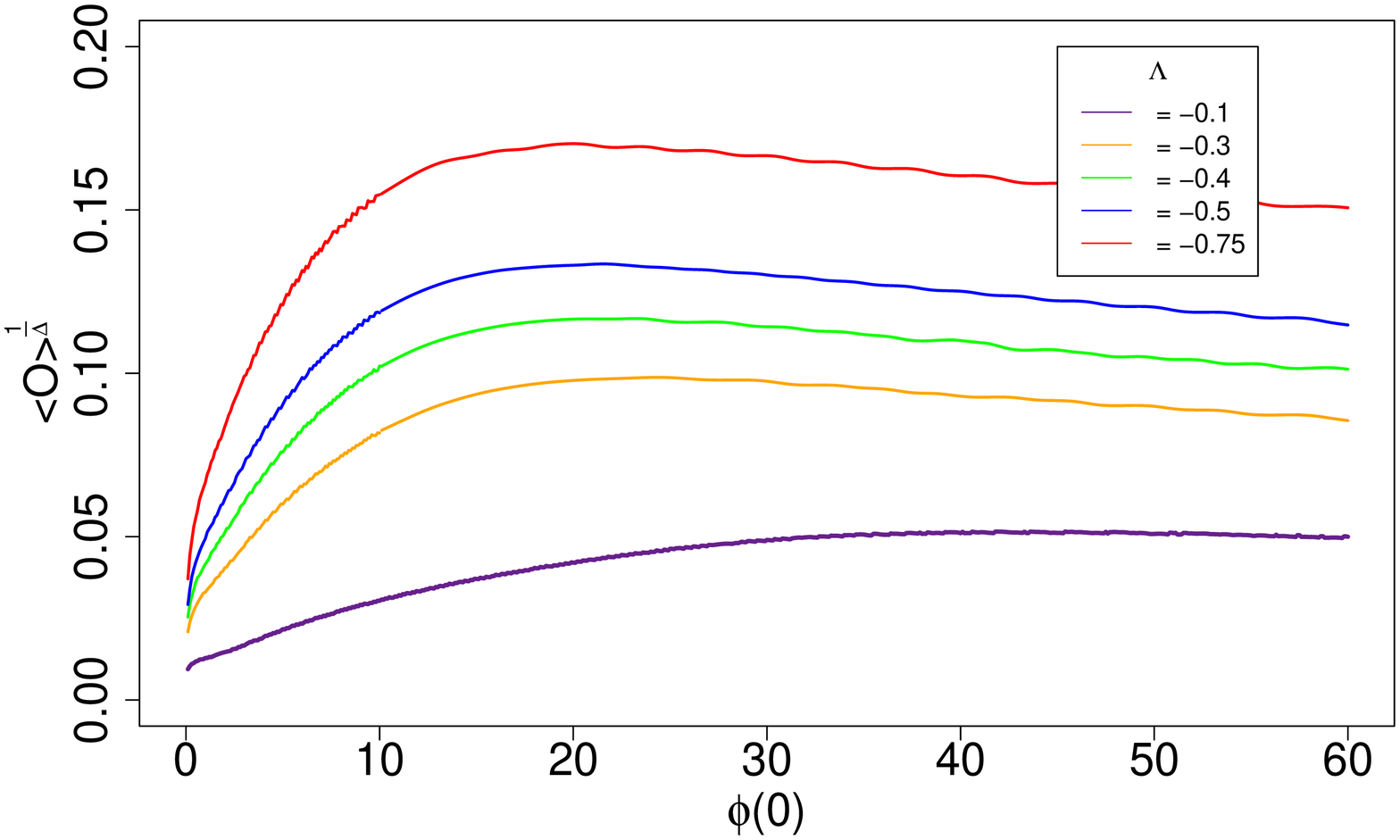}}\\
\subfigure[][$<O>^{\frac{1}{\Delta}}$ over charge $Q$]{\label{holo1_2}
\includegraphics[width=5.5cm,angle=270]{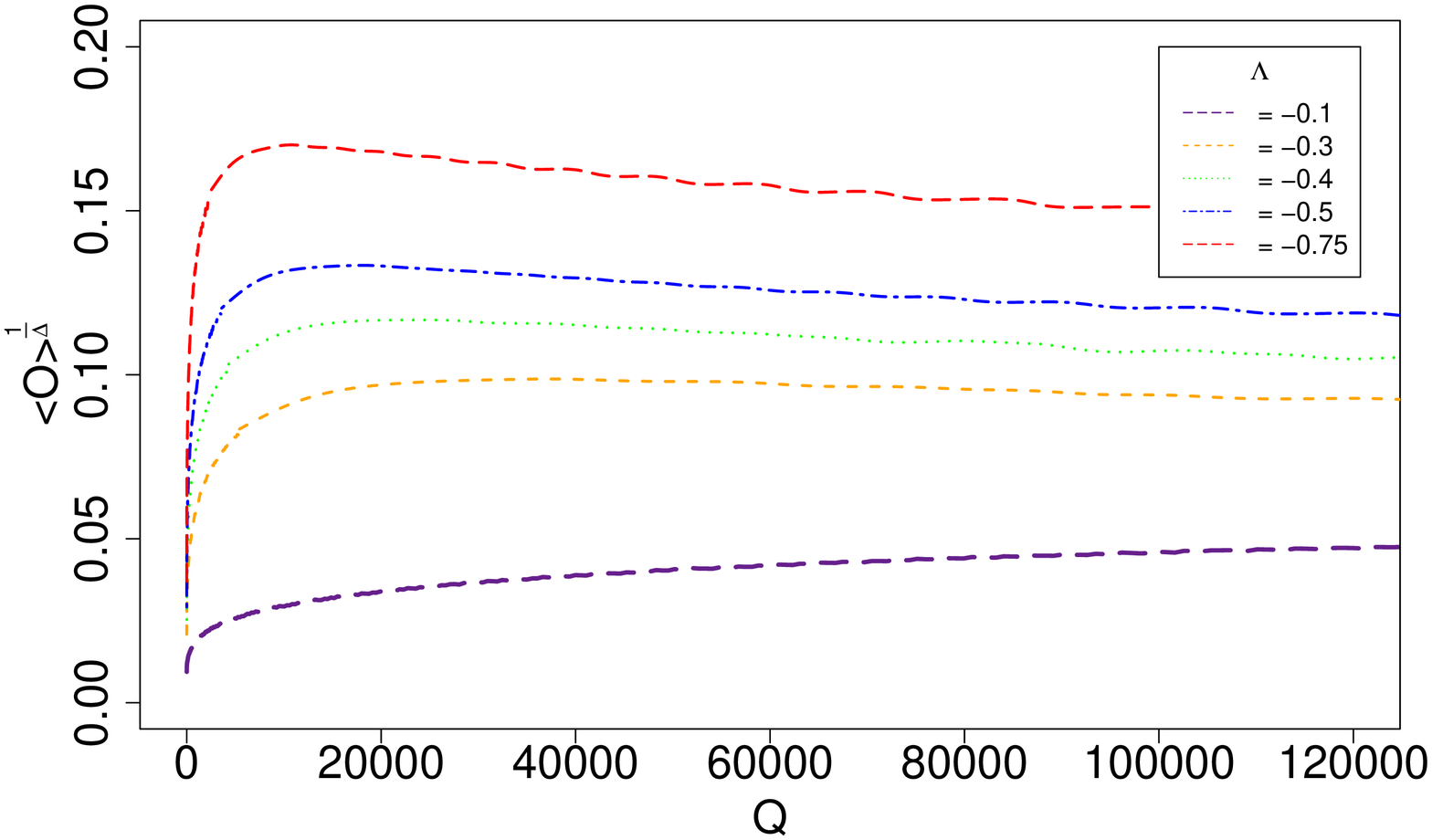}} 
\subfigure[][$<O>^{\frac{1}{\Delta}}$ over mass $M$ ]{\label{holo1_3}
\includegraphics[width=5.5cm,angle=270]{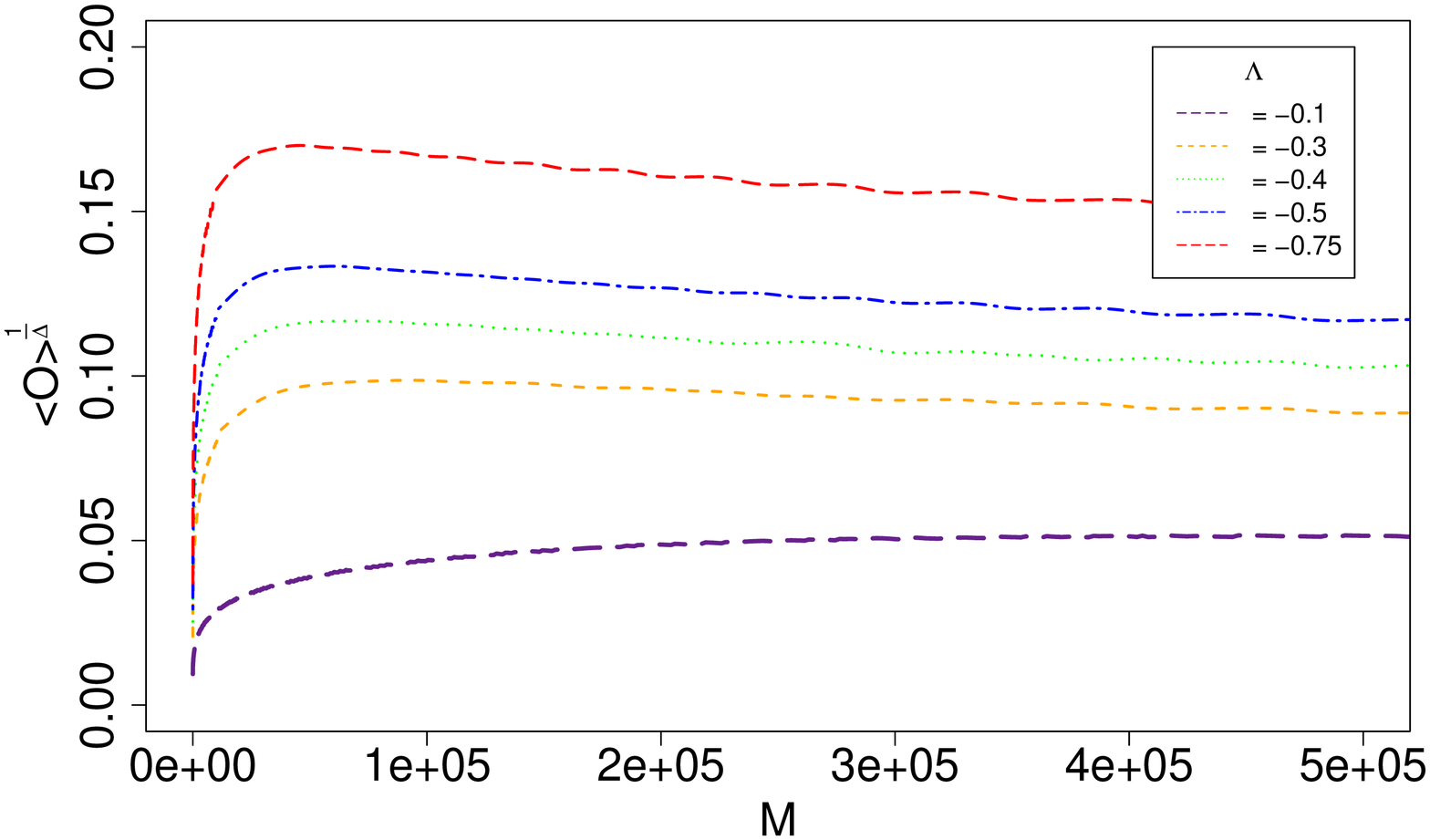}} 
\end{center}
\caption{\label{holo1} The expectation value of the dual operator on the boundary
$<O>^{\frac{1}{\Delta}}\equiv \phi_{\Delta}^{\frac{1}{\Delta}}$ as function of $\phi(0)$ (a), as function
of the charge $Q$ (b) and as function of the mass $M$ (c), respectively, for $Q$-balls in
an Anti-de Sitter background with different values of the cosmological constant $\Lambda$.  }
\end{figure}

\begin{figure}[h]
\begin{center}

\subfigure[][$<O>^{\frac{1}{\Delta}}$ over $\phi(0)$]{\label{holo2_1}
\includegraphics[width=6cm,angle=270]{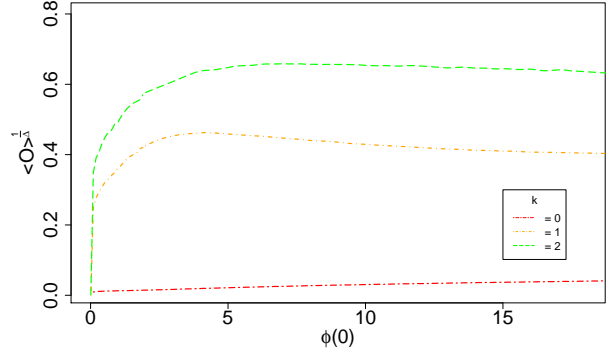}}\\
\subfigure[][$<O>^{\frac{1}{\Delta}}$ over charge $Q$]{\label{holo2_2}
\includegraphics[width=5.5cm,angle=270]{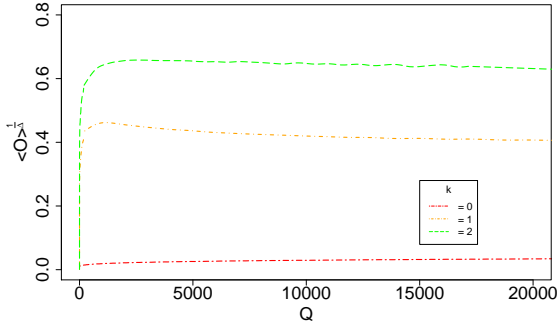}} 
\subfigure[][$<O>^{\frac{1}{\Delta}}$ over mass $M$ ]{\label{holo2_3}
\includegraphics[width=5.5cm,angle=270]{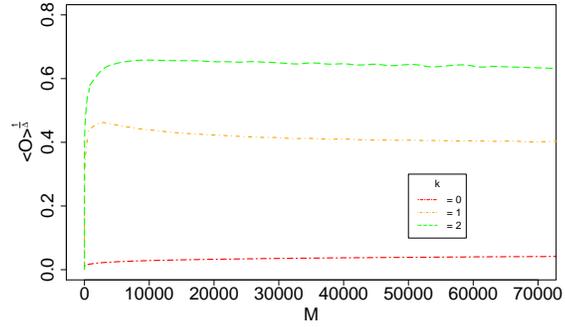}} 
\end{center}
\caption{\label{holo_ex1} The expectation value of the dual operator on the boundary
$<O>^{\frac{1}{\Delta}}\equiv \phi_{\Delta}^{\frac{1}{\Delta}}$ as function of $\phi(0)$ (a), as function
of the charge $Q$ (b) and as function of the mass $M$ (c), respectively, for fundamental ($k=0$)
and radially excited $Q$-balls ($k=1,2$) in an Anti-de-Sitter background with $\Lambda=-0.1$. }
\end{figure}

\begin{figure}[h]
\begin{center}

\subfigure[][$<O>^{\frac{1}{\Delta}}$ over $\phi(0)$]{\label{holo2_1}
\includegraphics[width=6cm,angle=270]{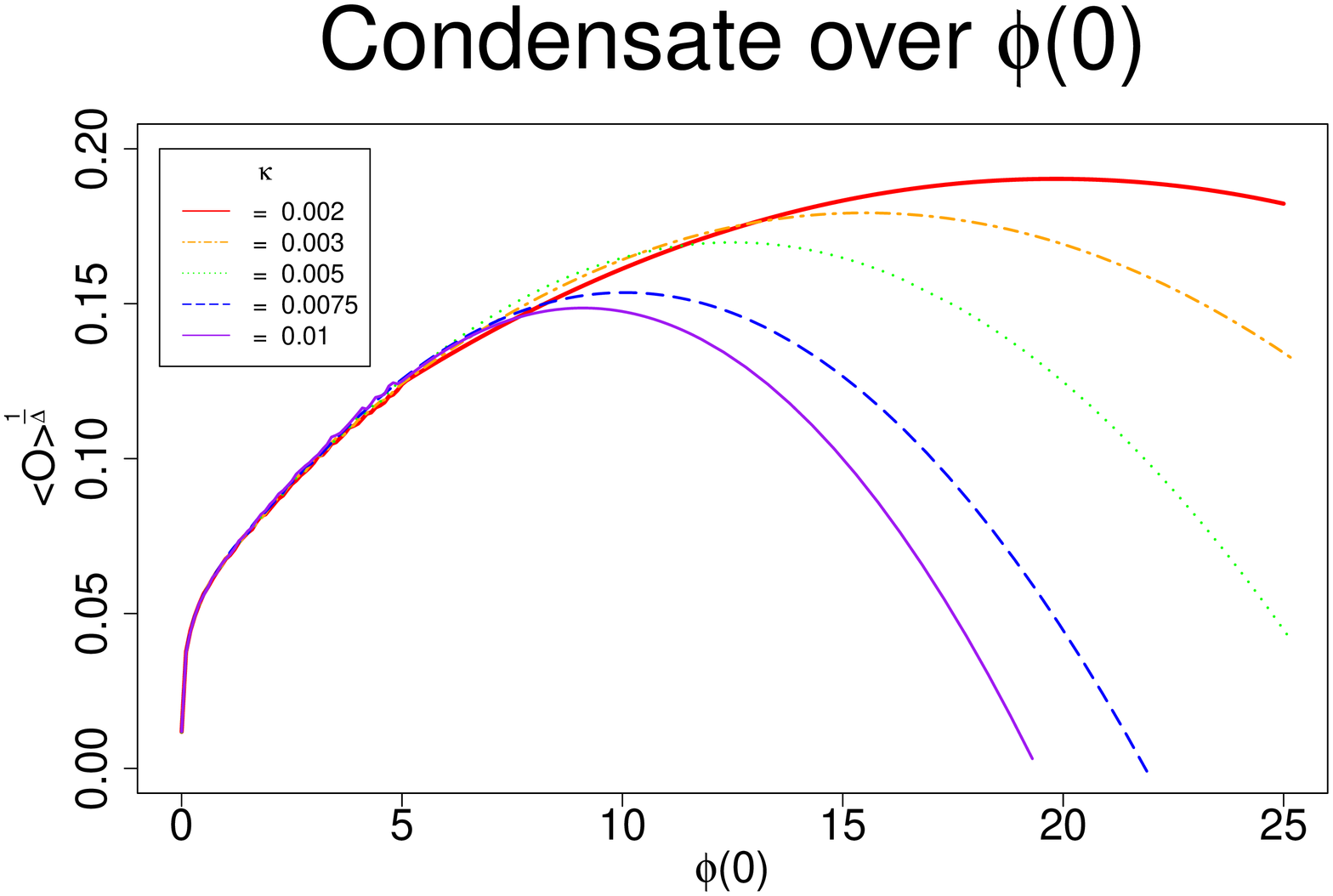}}\\
\subfigure[][$<O>^{\frac{1}{\Delta}}$ over charge $Q$]{\label{holo2_2}
\includegraphics[width=5.5cm,angle=270]{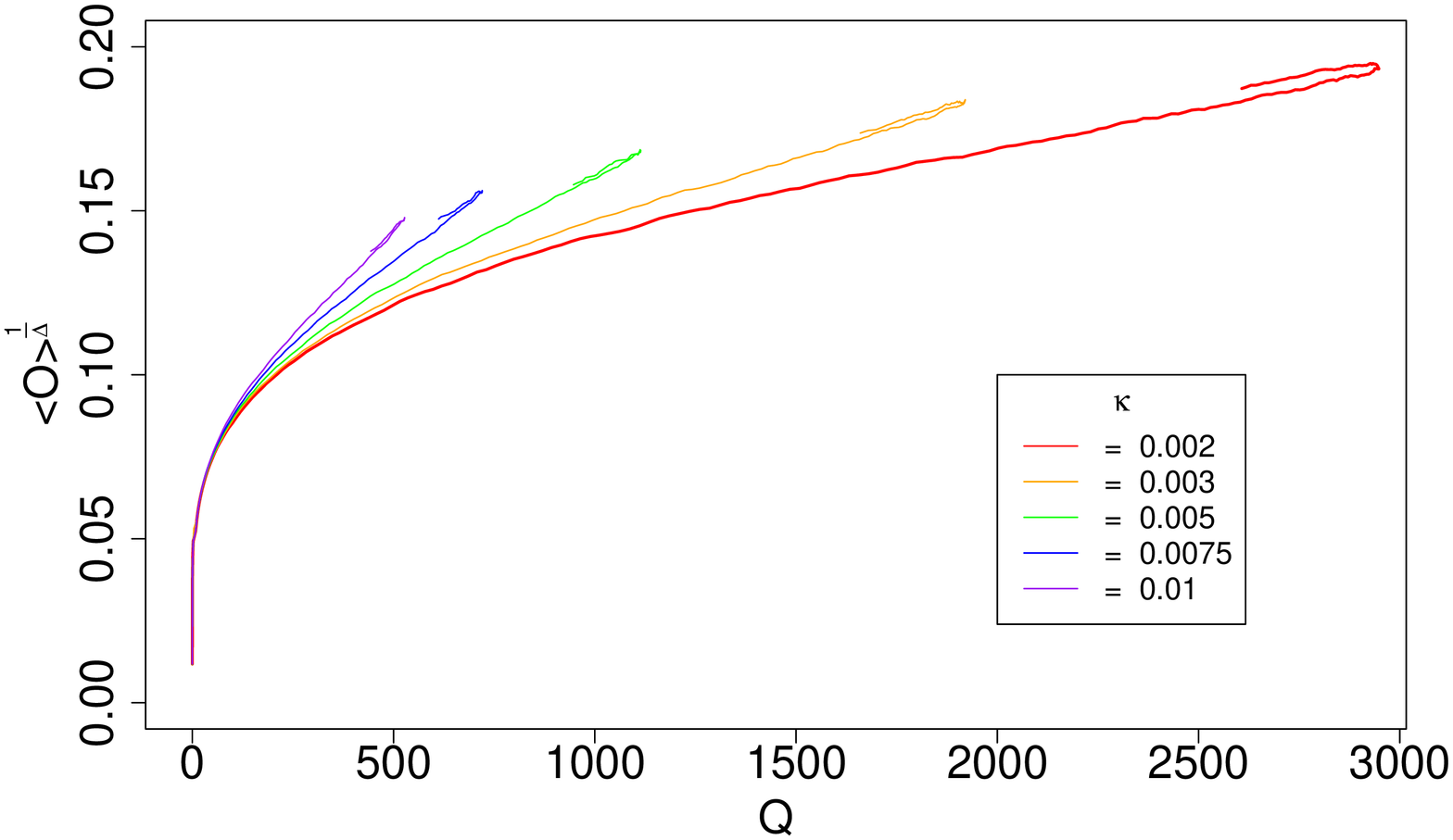}} 
\subfigure[][$<O>^{\frac{1}{\Delta}}$ over mass $M$ ]{\label{holo2_3}
\includegraphics[width=5.5cm,angle=270]{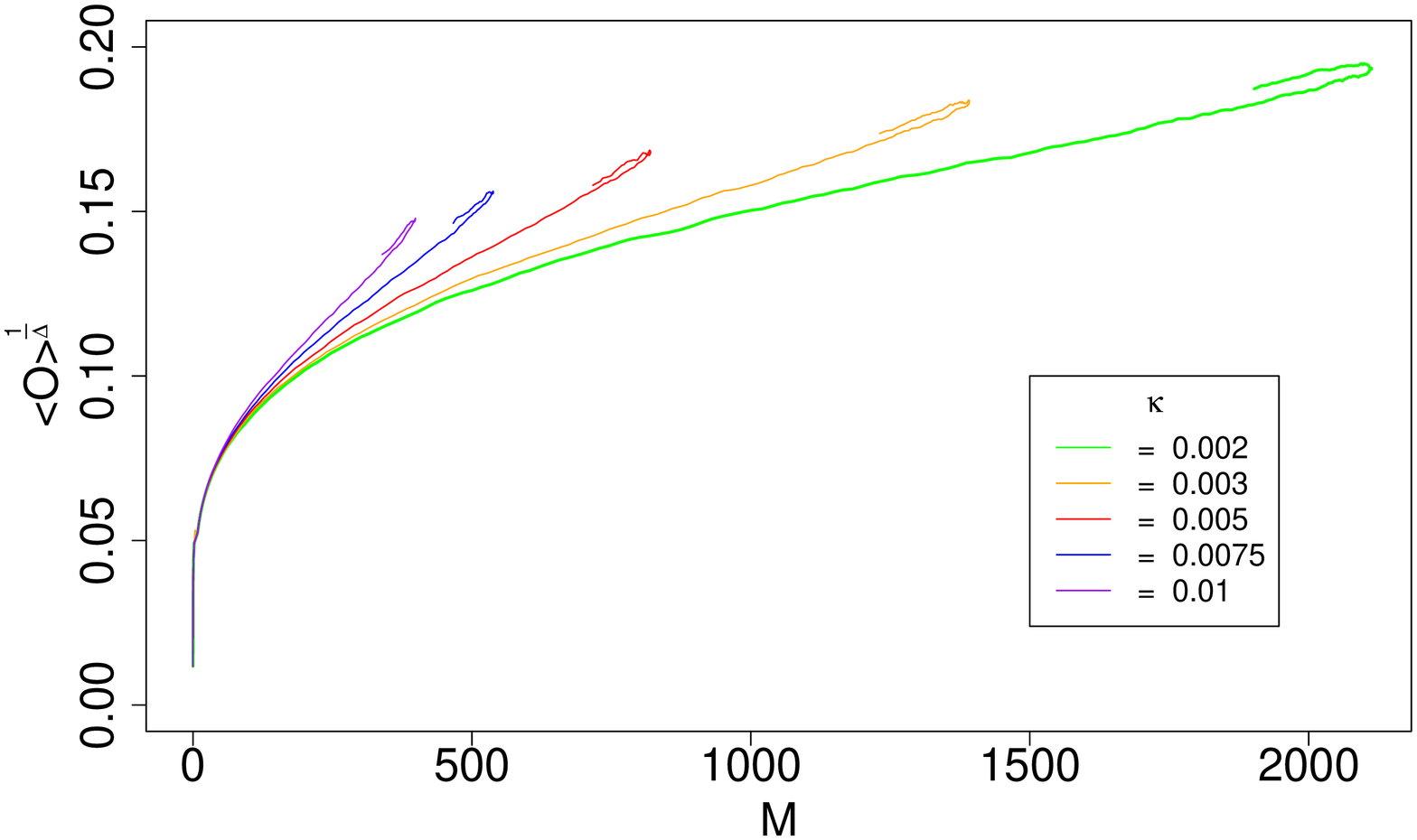}} 
\end{center}
\caption{\label{holo2} The expectation value of the dual operator on the boundary
$<O>^{\frac{1}{\Delta}}\equiv \phi_{\Delta}^{\frac{1}{\Delta}}$ as function of $\phi(0)$ (a), as function
of the charge $Q$ (b) and as function of the mass $M$ (c), respectively, for boson stars
in Anti-de Sitter space-time for different values of $\kappa$ and $\Lambda=-0.75$.  }
\end{figure}

\begin{figure}[h]
\begin{center}

\subfigure[][$<O>^{\frac{1}{\Delta}}$ over $\phi(0)$]{\label{holo2_1}
\includegraphics[width=6cm,angle=270]{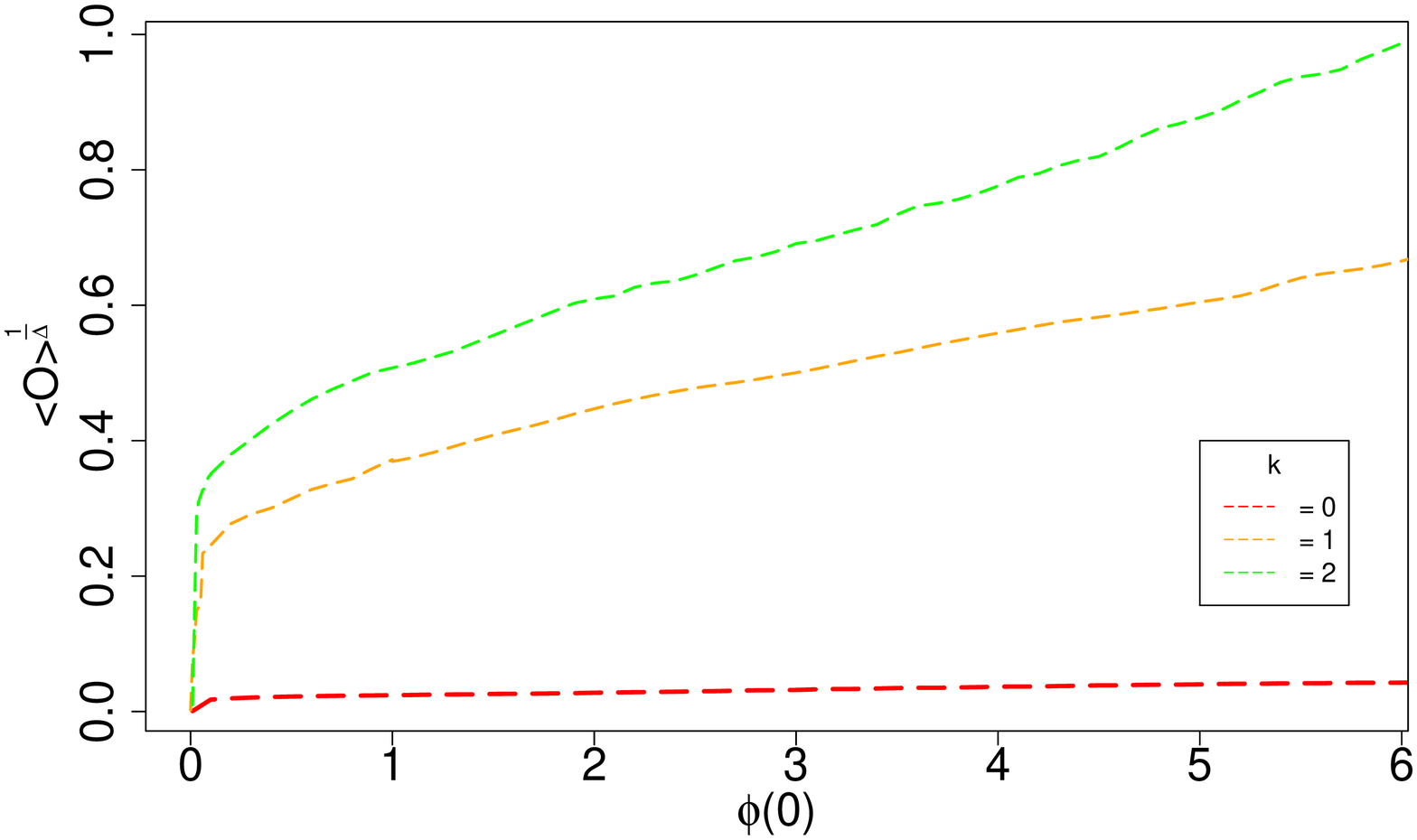}}\\
\subfigure[][$<O>^{\frac{1}{\Delta}}$ over charge $Q$]{\label{holo2_2}
\includegraphics[width=5.5cm,angle=270]{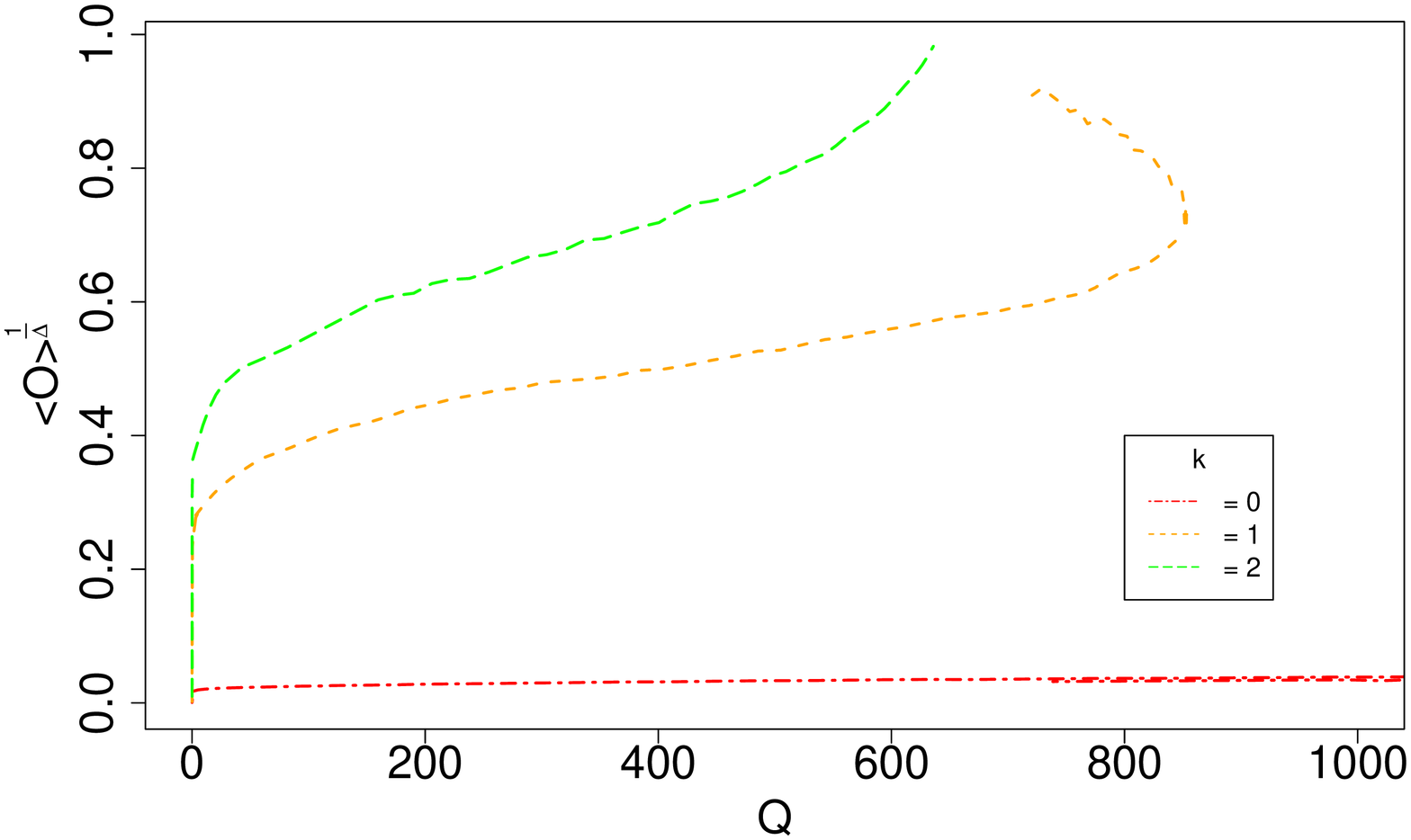}} 
\subfigure[][$<O>^{\frac{1}{\Delta}}$ over mass $M$ ]{\label{holo2_3}
\includegraphics[width=5.5cm,angle=270]{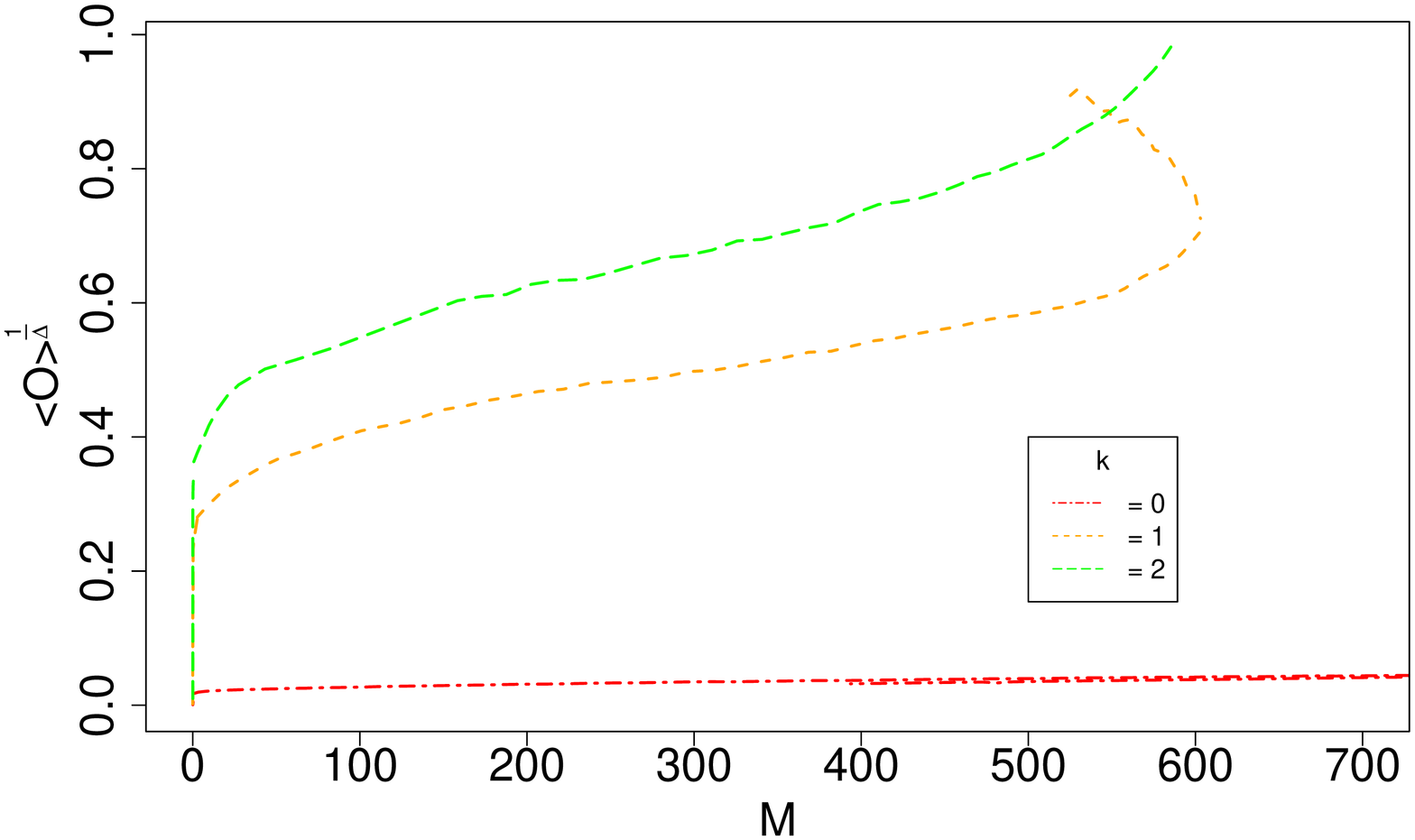}} 
\end{center}
\caption{\label{holo2ex} The expectation value of the dual operator on the boundary
$<O>^{\frac{1}{\Delta}}\equiv \phi_{\Delta}^{\frac{1}{\Delta}}$ as function of $\phi(0)$ (a), as function
of the charge $Q$ (b) and as function of the mass $M$ (c), respectively, for boson stars
in Anti-de Sitter space-time for $\kappa=0.01$, $\Lambda=-0.1$ and $k=0,1,2$, respectively.  }
\end{figure}

\section{Holographic interpretation: condensate of glueballs}
In the following, we want to use the AdS/CFT correspondence to interpret our solutions in AdS space-time
in terms of a Bose-Einstein condensate of scalar glueballs. This has been suggested in \cite{horowitz} 
for planar boson star solutions. Here, we are studying global AdS and hence our dual and strongly coupled
CFT lives on the surface of a 2-sphere. 

\subsection{Probe limit}
We have first studied the case $\kappa=0$, i.e. neglecting the backreaction of the space-time on
the condensate of glueballs. In Fig.\ref{holo1} we show the value of the
condensate $<O>^{1/\Delta}\equiv \phi_{\Delta}^{1/\Delta}$ in dependence on $\phi(0)$ (a), 
in dependence on the charge $Q$ (b) and in dependence on the mass $M$ (c), respectively.

We find that the condensate is an increasing function of $\phi(0)$, $Q$ and $M$, respectively.
Moreover, decreasing the cosmological constant increases the value of the condensate for a fixed value
of $\phi(0)$, $Q$ and $M$. This is related to the fact that the increase of the absolute value of
the negative cosmological constant is, of course, related to a decrease of the radius
of the AdS space-time. This in turn allows to built up larger condensates. 
We find that at some intermediate values of $\phi(0)$, $Q$ and $M$ the condensate has a maximal
value. Our numerical results suggest that the larger the absolute value of $\Lambda$ the smaller
in the value of $\phi(0)$ at which the condensate reaches its maximum. However, the corresponding
values of the mass and charge at which the maximal condensate appears are significantly larger than the
values of the mass and charge necessary to form a condensate in the first place.

We have also studied the value of the condensate for the radially excited $Q$-ball solutions mentioned
above. Our results are shown in Fig.\ref{holo_ex1}, where we present $<O>^{1/\Delta}\equiv \phi_{\Delta}^{1/\Delta}$ in dependence on $\phi(0)$ (a), 
in dependence on the charge $Q$ (b) and in dependence on the mass $M$ (c), respectively, in an AdS background with $\Lambda=-0.75$
and for the fundamental solution without nodes ($k=0$) and for the radially excited solutions with $k=1,2$ nodes.
It is apparent that the value of the condensate increases with increasing node number for fixed $\phi(0)$, $M$ or $Q$.
Hence, larger condensates can built up in the presence of radially excited $Q$-ball solutions. However, since
we expect these radially excited solutions to be unstable to decay to the fundamental solution, these condensates
are very likely unstable. It would have to be computed explicitely how long-lived these excited solutions
are - and hence of how much physical relevance these condensates are.

\subsection{Including backreaction}
This corresponds to $\kappa\neq 0$. Our results for fixed $\Lambda=-0.75$ and different
values of the backreaction are shown in Fig.\ref{holo2}.
Plotting the condensate as function of $\phi(0)$ (see Fig.\ref{holo2_1}) we find that a value of 
$\phi(0)=\phi(0)_{\rm max}$ exists at which the condensate has a maximal value. 
This was also observed for planar boson stars in \cite{horowitz}. We find that the larger the backreaction, i.e.
the larger $\kappa$ the smaller is $\phi(0)_{\rm max}$.

Moreover, at a critical value of $\phi(0)=\phi(0)_{\rm crit} > 0$, the condensate tends to zero again.
Hence, glueball condensates are only possible in a finite range of the parameter $\phi(0)$. The interval
of $\phi(0)$ on which glueball condensates exist decreases when increasing the backreaction.

When plotting the condensate as function of the charge $Q$ and the mass $M$ (see Fig.\ref{holo2_2} and Fig.\ref{holo2_3}, respectively)
we find that glueball condensates exist only up to a maximal finite mass of the charge and mass.
Moreover, the value of the condensate increases for fixed $Q$ or $M$ when increasing the backreaction.
This is a standard observation in applications of the AdS/CFT correspondence applied to strongly coupled
QFTs (see e.g. \cite{hhh} and references therein). Interestingly, we observe that in a small interval close to the maximal charge $Q$ (respectively maximal
mass $M$) two different values of the condensate are possible for a fixed $M$ and $Q$. One of the branches corresponds
to a stable boson star, while the other would probably be unstable. 
For the unstable boson star, the condensate can be larger. However, it would likely be much shorter lived than that
building up for a stable boson star. In order to decide how short-lived this is, one would probably
have to simulate the dynamical decay of the unstable boson star to the stable one. This is out of the scope of the
present paper.

We have also studied the condensate building up on radially excited boson stars in AdS space-time. Our results
for $\kappa=0.01$ and $\Lambda=-0.1$ 
are given in Fig.\ref{holo2ex}. We show the condensate for the fundamental solution ($k=0$) as well
as for the two radially excited solutions with $k=1,2$. It is apparent that the values of the
condensate can be much higher in the space-time of radially excited boson stars. This is similar to
what we have observed for the $Q$-balls in an AdS background (see above). Moreover, we find that the
difference of the condensate on the first and second branch of solutions increases with increasing node number.
While the difference between the two branches for $k=0$ is quite small, it is already quite big for $k=1$. Again,
we expect these solutions to be unstable and hence the condensates to decay.

\section{Summary and discussions}
In this paper, we have studied AdS $Q$-balls and boson stars that arise in a complex
scalar field model with exponential potential. This potential appears in gauge-mediated SUSY breaking
in the MSSM. We find that the qualitative features in the case of an asymptotically flat space-time are similar
to those observed for other potentials. When choosing a non-vanishing cosmological constant
the qualitative features change. A study on boson stars in AdS already exists for a massive
scalar field without self-interaction \cite{radu}, however, in this latter paper the frequency has been scaled to one.
In the present paper, we have kept the frequency as a parameter and have studied the dependence of the
mass and charge of the solutions on this parameter. We observe that $Q$-balls in an AdS background
exist for arbitrary values of the charge and mass and that only one branch of solutions exists. 
Moreover, we find that radially excited solutions exist both for a Minkowski as well as for an AdS background.
When including backreaction of the space-time, we find that boson stars exist only up to a maximal
value of the mass and charge and that at the minimal value of the frequency a spiraling behaviour appears.
Plotting the value of the scalar field on the AdS boundary and interpreting our solutions as holographic
duals of Bose-Einstein condensates of scalar glueballs, we find that a maximal value of the condensate
exists in the probe limit as well as in the backreacted case. Moreover, for fixed values of the charge
and mass of the solutions, we find that more than one value of the condensate is possible. However, the condensate
with the larger value is (likely) associated to an unstable solution and would hence decay. 

In this paper, we have only studied non-spinning $Q$-balls and boson stars. It would be interesting
to extend our results to the case of rotation. These solutions have already been studied for a $\phi^6$-potential
in asymptotically flat space-time \cite{kk1,kk2} and we would expect that they also exist in asymptotically AdS and with
an exponential potential.
The question is then whether the boundary theory would also possess rotation and if one could
hence describe rotating Bose-Einstein condensates of glueballs with our model. This is currently under investigation.\\
\\

{\bf Acknowledgement} We gratefully acknowledge the Deutsche Forschungsgemeinschaft (DFG) for financial support
within the framework of the DFG Research Training group 1620 {\it Models of gravity}.

\end{document}